\documentclass[aps,twocolumn,superscriptaddress,english,10pt,nofootinbib]{revtex4-2}

\usepackage{float}

\usepackage{amssymb, amsmath, bm, dcolumn, epsf, graphicx, latexsym, slashed, simplewick}
\usepackage[utf8]{inputenc}

\usepackage{color}

\def\be{\begin{equation}}
\def\ee{\end{equation}}
\def\bea{\begin{eqnarray}}
\def\eea{\end{eqnarray}}

\usepackage{hyperref}
\usepackage{comment}

\usepackage[normalem]{ulem}

\begin{document}

\title{Constraining Early Dark Energy with Large-Scale Structure}

\author{Mikhail M. Ivanov}
\affiliation{Center for Cosmology and Particle Physics, Department of Physics, New York University,\\ New York, NY 10003, USA}
\affiliation{Institute for Nuclear Research of the Russian Academy of Sciences,\\ 60th October Anniversary Prospect, 7a, 117312 Moscow, Russia}

\author{Evan McDonough}
\affiliation{Center for Theoretical Physics, Massachusetts Institute of Technology,\\ Cambridge, MA 02139, USA}

\author{J.~Colin Hill}
\affiliation{Department of Physics, Columbia University, New York, NY, USA 10027}
\affiliation{Center for Computational Astrophysics, Flatiron Institute, New York, NY, USA 10010}

\author{Marko Simonovi\'{c}}
\affiliation{Theoretical Physics Department, CERN,\\ 1 Esplanade des Particules, Geneva 23, CH-1211, Switzerland}

\author{Michael W.~Toomey}
\affiliation{Brown Theoretical Physics Center and Department of Physics,\\ Brown University, Providence, RI 02912, USA}

\author{Stephon Alexander}
\affiliation{Brown Theoretical Physics Center and Department of Physics,\\ Brown University, Providence, RI 02912, USA}

\author{Matias Zaldarriaga}
\affiliation{School of Natural Sciences, Institute for Advanced Study,\\ 1 Einstein Drive, Princeton, NJ 08540, USA}

\begin{abstract}
An axion-like field comprising $\sim 10\%$ of the energy density of the universe near matter-radiation equality is a candidate to resolve the Hubble tension; this is the ``early dark energy'' (EDE) model. However, as shown in~Hill~et~al.~(2020)~\cite{Hill:2020osr}, the model fails to simultaneously resolve the Hubble tension and maintain a good fit to both cosmic microwave background (CMB) and large-scale structure (LSS) data. Here, we use redshift-space galaxy clustering data to sharpen constraints on the EDE model. We perform the first EDE analysis using the full-shape power spectrum likelihood from the Baryon Oscillation Spectroscopic Survey (BOSS), based on the effective field theory (EFT) of LSS.  The inclusion of this likelihood in the EDE analysis yields a $25\%$ tighter error bar on $H_0$ compared to primary CMB data alone, yielding $H_0=68.54^{+0.52}_{-0.95}$ km/s/Mpc~($68\%$~CL).  In addition, we constrain the maximum fractional energy density contribution of the EDE to $f_{\rm EDE}<0.072$ ($95\%$~CL).  We explicitly demonstrate that the EFT BOSS likelihood yields much stronger constraints on EDE than the standard BOSS likelihood. Including further information from photometric LSS surveys,the constraints narrow by an additional $20\%$, yielding $H_0=68.73^{+0.42}_{-0.69}$~km/s/Mpc~($68\%$~CL) and $f_{\rm EDE}<0.053$~($95\%$~CL). These bounds are obtained without including local-universe $H_0$ data, which is in strong tension with the CMB and LSS, even in the EDE model.  We also refute claims that MCMC analyses of EDE that omit SH0ES from the combined dataset yield misleading posteriors.  Finally, we show that upcoming {\it Euclid}/DESI-like spectroscopic galaxy surveys will greatly improve the EDE constraints. We conclude that current data preclude the EDE model as a resolution of the Hubble tension, and that future LSS surveys can close the remaining parameter space of this model.
\end{abstract}

\maketitle

\section{Introduction}

The persistent and growing discrepancy in the value of the Hubble constant, $H_0$, inferred from different observations~\cite{Verde:2019ivm}, if taken at face value, presents a serious challenge to the standard cosmological model.  This tension is conventionally viewed as that between the value inferred from the cosmic microwave background (CMB)~\cite{Aghanim:2018eyx} and the SH0ES measurement of the classical distance ladder utilizing Type Ia supernovae (SNIa) at cosmological redshifts~\cite{Riess:2019cxk}. Indeed, the {\it Planck} 2018 CMB and the SH0ES measurements of $H_0$ disagree at $4.4 \sigma$ statistical significance, with the two given by $H_0=67.37 \pm 0.54$ km/s/Mpc and $H_0=74.03 \pm 1.42\,{\rm km/s /Mpc}$ for {\it Planck}~\cite{Aghanim:2018eyx} and SH0ES~\cite{Riess:2019cxk}, respectively.
More recently it has become apparent that the $H_0$ tension is not restricted to the CMB vs.~SH0ES, but on the contrary, ranges from 4-6$\sigma$ significance for a large array of differing data sets and data set combinations \cite{Verde:2019ivm}, which are commonly split in two categories: early-universe and late-universe measurements.\footnote{This nomenclature can be somewhat misleading, since some of the measurements of $H_0$ that entirely rely on the late-universe datasets would be classified as early-universe measurements. A more meaningful distinction is between the {\em direct} (or local) measurements of $H_0$ such as SH0ES, versus {\em indirect} (or global) measurements obtained from a fit of $\Lambda$CDM or any other model to cosmological data such as the CMB or galaxy clustering, regardless of the redshift of these datasets.}

There are several local or late-universe measurements that all lead to values of $H_0$ consistent with SH0ES. For instance, the Cepheids utilized for calibration of the cosmic distance ladder by SH0ES can be replaced with Miras, variable red giant stars, leading to $H_0=73.3\pm3.9\,{\rm km/s /Mpc}$ \cite{Huang:2019yhh}, or traded for the ``tip of the red giant branch'' in the Hertzsprung-Russell diagram, which yields a somewhat lower value $H_0 = 69.6 \pm 1.9 \,{\rm km/s /Mpc}$ \cite{Freedman:2020dne}. Alternatively, local measurements can be performed in a variety of ways that are independent of the cosmic distance ladder, e.g., through the measurement of time delays in strongly lensed quasar systems, which yields $H_0= 73.3^{+1.7} _{-1.8}\, {\rm km/s /Mpc}$ \cite{Wong:2019kwg},\footnote{However, some concerns about unaccounted systematics in lensing time-delay measurements have been raised, for instance in~\cite{Kochanek:2019ruu,Blum:2020mgu}; indeed, the latest strong-lensing analyses, after accounting for these systematics, have now found a lower value of $H_0$ with significantly increased error bars, $H_0 = 67.4^{+4.1}_{-3.2}$ km/s/Mpc~\cite{Birrer2020}.} through very-long-baseline interferometry observations of water masers orbiting supermassive black holes, which yields $H_0=73.9\pm3.0\, {\rm km/s /Mpc}$ \cite{Pesce:2020xfe}, and through gravitational waves from merging binary neutron stars \cite{Abbott:2017xzu, Soares-Santos:2019irc}.

On the other side, there are several cosmological measurements of $H_0$ that can be made at low redshifts and independently of the CMB anisotropy data. This is done by combining various large-scale structure (LSS) observations with a prior on the baryon density $\omega_{\rm b}$ inferred from Big Bang nucleosynthesis (BBN) \cite{2016ApJ...830..148C}. One such measurement comes from the baryon acoustic oscillation (BAO) experiments, such as the Baryon Oscillation Spectroscopic Survey (BOSS). Measurements of the BAO scale for galaxies and the Ly$\alpha$ forest \cite{Blomqvist:2019rah}, together with the BBN prior, lead to $H_0 =67.6\pm1.1 $ km/s/Mpc \cite{Aubourg:2014yra,Cuceu:2019for,Schoneberg:2019wmt}.  Similarly, the combination of Dark Energy Survey (DES) \cite{Abbott:2017wau} data and BOSS BAO data gives $H_0 = 67.4 ^{+1.1} _{-1.2}\, {\rm km/s /Mpc}$ \cite{Abbott:2017smn}. Measurements of the Hubble constant from galaxy clustering alone (without the Ly$\alpha$ data) can also be done using the full shape (FS) of the galaxy power spectrum  \cite{Ivanov:2019pdj,DAmico:2019fhj,2020A&A...633L..10T,Philcox:2020vvt}. In particular, the joint FS+BAO data from BOSS yields $H_0 = 68.6 \pm 1.1$ km/s/Mpc \cite{Philcox:2020vvt}, in excellent agreement with the CMB result.

In all these measurements, standard early-universe physics is assumed, such that the sound horizon $r_s$ is a fixed function of the $\Lambda$CDM cosmological parameters. On the other hand, the angular scale of the sound horizon, $\theta_s$, is measured to a very high precision of $0.03\%$ by the CMB data~\cite{Aghanim:2018eyx}. This renders any sizeable shift in $H_0$, such as that necessary to resolve the tension with SH0ES, incompatible with data, unless new physics is introduced in the early universe so as to change $r_s$ (and $H_0$) while keeping the angular scale $\theta_s$ fixed. Such early universe solutions to the tension have been advocated in, e.g.,~\cite{Knox:2019rjx}; a prototypical model realization goes by the name ``early dark energy'' (EDE)~\cite{Poulin:2018cxd}. Many EDE-like models have been proposed, both in the context of the $H_0$ tension \cite{Poulin:2018cxd,Smith:2019ihp,Agrawal:2019lmo,Alexander:2019rsc,Lin:2019qug,Sakstein:2019fmf,Niedermann:2019olb,Kaloper:2019lpl,Berghaus:2019cls} and other areas of cosmological phenomenology~(e.g.,~\cite{Kamionkowski:2014zda,Poulin:2018a,Hill2018}).

In the EDE scenario one postulates an additional dynamical scalar field, which behaves like dark energy until a critical time near matter-radiation equality, at which point its energy density rapidly decays. The increased energy density at early times serves to decrease the comoving sound horizon, such that an increased $H_0$ can be accommodated while keeping $\theta_s$ fixed. This comprises a 3-parameter extension to $\Lambda$CDM, defined by a critical redshift $z_c$, the peak EDE energy density fraction of the universe $f_{\rm EDE}$, and the initial value of the scalar field, denoted by the dimensionless quantity $\theta_i$ (analogous to the axion misalignment angle \cite{Dine:1982ah,Abbott:1982af,Preskill:1982cy}).\footnote{An additional fourth parameter, $n$, denoting an exponent of the scalar field potential $V\propto (1 -\cos \theta)^n$, can be included, and is only weakly constrained; the best-fit integer value is $n=3$~\cite{Smith:2019ihp}.} Remarkably, the EDE model allows for values of $H_0$ in near-agreement with SH0ES whilst leaving the fit to the CMB spectra nearly unchanged from that in $\Lambda$CDM.

However, the EDE scenario begins to falter when confronted with LSS data~\cite{Hill:2020osr}. While the EDE field does not directly impact the formation of structure at late times (due to its rapid decay around matter-radiation equality), the accompanying shifts in the standard $\Lambda$CDM parameters, necessary to retain the fit to CMB data, become clearly detectable thanks to the breaking of various degeneracies when combining CMB and LSS data. In addition to CMB lensing, BOSS BAO, and BOSS redshift-space distortion (RSD) data, the analysis of~\cite{Hill:2020osr} included the DES-Y1 data \cite{Abbott:2017wau} and the weak gravitational lensing measurements from KiDS+VIKING-450 (KV-450)~\cite{Hildebrandt:2016iqg,2020A&A...633A..69H} and the Subaru Hyper Suprime-Cam (HSC)~\cite{Hikage:2018qbn}. If these additional LSS data are included, the evidence for EDE is below 2$\sigma$, even when SH0ES is included in the analysis~\cite{Hill:2020osr}. This indicates that the SH0ES measurement remains an outlier in the EDE scenario, just as in $\Lambda$CDM. Indeed, if SH0ES is removed from the combined data set and all LSS data is included (``walking barefoot'', Sec.~VI.E of \cite{Hill:2020osr}), one finds an upper bound $f_{\rm EDE}<0.053$ at 95\% CL, well below the value claimed to resolve the Hubble tension, $f_{\rm EDE} \simeq 0.107$ \cite{Smith:2019ihp}.

One caveat behind this result is that the ``compressed'' BOSS RSD likelihood used in the ``walking barefoot'' analysis of~\cite{Hill:2020osr} was derived implicitly assuming standard early-universe physics and {\em fixed-shape} template for the galaxy power spectrum.  In the official BOSS data analysis this is implemented through the so-called ``shape priors", fixing the physical cold dark matter and baryon densities $\omega_{\rm cdm}$ and $\omega_{\rm b}$ to the best-fit {\it Planck} values obtained from the cosmological analysis within $\Lambda$CDM~\cite{Alam:2016hwk,Beutler:2016arn}. 
Even though this method is referred to in the literature as the full-shape BOSS analysis, we stress that no power spectrum shape information is used in this procedure. Therefore, one may wonder to what extent the use of such a $\Lambda$CDM-based and fixed-shape likelihood may have impacted the conclusions of previous analyses for the EDE model~\cite{Poulin:2018cxd,Agrawal:2019lmo,Smith:2019ihp,Hill:2020osr,Niedermann:2020dwg}.
Addressing this question is one of the main goals of this paper.\footnote{One may be concerned that
some implicit early-universe assumptions may also impact the BAO measurements. However, in contrast to 
the RSD, the BAO measurements by construction are largely unaffected by assumptions about cosmology. First, the use of the Alcock-Paczynski scaling parameters \cite{Alcock:1979mp} to extract the BAO frequency from the 3-dimensional galaxy distribution is accurate if the Universe has a local Friedman-Robertson-Walker geometry (see, e.g., Refs.~\cite{Matsubara:1996nf,Ballinger:1996cd,Heinesen:2019phg}), which clearly holds true in the EDE model.  Second, the BAO frequency can be extracted from the post-reconstruction position space correlation function (or the power spectrum) by fitting the BAO peak (or Fourier-space BAO wiggles) with a simple Gaussian (or harmonic) template, which does not rely on any cosmology-specific information. Thus,
the BAO measurements would not be biased even if a
reasonable cosmology-dependent fiducial template were used~\cite{Bernal:2020vbb}.} 

To that end, we repeat the analysis of~\cite{Hill:2020osr} using a new BOSS likelihood which: (a) is tailored for the EDE model (i.e.,~no implicit $\Lambda$CDM-based assumptions are made), (b) has {\em all} relevant cosmological parameters varied, and (c) uses the {\em full shape} of the redshift-space galaxy power spectrum, 
going beyond the simplified
$f\sigma_8+$BAO parametrization. 
This analysis has been made possible by virtue of recent progress in LSS theory.
First, the consistent formulation 
of perturbation theory has been finalized in the form of the Effective Field Theory (EFT) of LSS (see~\cite{DAmico:2019fhj,Ivanov:2019pdj} and references therein). 
Various ingredients of this approach (e.g.,~UV counterterms and IR resummation)
have been independently derived in many different setups.
Second, there was significant 
improvement in numerical methods,
which allowed one to build
extensions of standard Boltzmann codes that consistently calculate the nonlinear galaxy clustering observables as a function of cosmological parameters~\cite{DAmico:2020kxu,Chudaykin:2020aoj}.\footnote{These tools have already been successfully used to constrain the base $\Lambda$CDM model using the BOSS data~\cite{Ivanov:2019pdj,DAmico:2019fhj}, 
verified in a blind simulation challenge~\cite{Nishimichi:2020tvu}, 
and applied to various extensions of the $\Lambda$CDM model, including massive neutrinos~\cite{Chudaykin:2019ock,Colas:2019ret,Ivanov:2019hqk}, varying number of relativistic degrees of freedom~\cite{Ivanov:2019hqk}, and dynamical dark energy~\cite{Ivanov:2019pdj,DAmico:2020kxu}.}
Exploiting all these results, we show that the shape information in the galaxy power spectrum is important and that the new BOSS likelihood used in our analysis indeed leads to much stronger constraints on EDE compared to the standard BOSS likelihood.

With these new results for the combined analysis of {\it Planck} and BOSS in hand, we turn to the question of how much additional information on EDE can be extracted from photometric weak lensing surveys. As justified in Ref.~\cite{Hill:2020osr}, we implement this extra information through a prior on $S_8\equiv \sigma_8 \left(\Omega_m/0.3 \right)^{0.5}$ and show that current data from DES, KV-450, and HSC tighten the upper bound on $f_{\rm EDE}$ even further, and lead to sharper constraints on $H_0$. Finally, with an eye towards {\it Euclid}~\cite{Amendola:2016saw} and the Dark Energy Spectroscopic Instrument (DESI) \cite{Levi:2019ggs}, we perform an EDE sensitivity forecast for upcoming spectroscopic galaxy surveys and show that they have the potential to definitively rule out the EDE model as a resolution to the Hubble tension.

The structure of this paper is as follows. In Sec.~II we review the EDE proposal, and in Sec.~III detail its imprint on LSS, in particular as it pertains to the BOSS data.  In Sec.~IV we present the constraints on EDE from current data, namely {\it Planck} 2018, BOSS full-shape+BAO, and an $S_8$ prior corresponding to the measurements of DES, KV-450, and  HSC. In Sec.~V we perform an EDE sensitivity forecast for {\it Euclid}. We close in Sec.~VI with a discussion of the implications for the $H_0$ tension, and directions for future work.  Additional results are collected in the appendices.

\section{The Early Dark Energy Proposal}
\label{sec:EDE}

The EDE scenario aims to increase the expansion rate in the early universe prior to recombination, while leaving the physics of the late universe unchanged. This is done in such a way as to not degrade the fit to CMB temperature and polarization data relative to $\Lambda$CDM. 

The increased expansion rate serves to reduce the comoving sound horizon at last scattering,
\begin{equation}
\label{eq:rs}
    r_s(z_*) = \int _{z_*} ^\infty \frac{{\rm d} z}{H(z)} c_s(z) ,
\end{equation}
where $z_*$ is the redshift of last scattering, such that an increased present-day expansion rate $H_0$, as encoded in the comoving angular diameter distance to last scattering,
\begin{equation}
\label{eq:DA}
  D_A(z_*) = \int _0 ^{z_*} \frac{{\rm d} z}{H(z)} ,
\end{equation}
can be accommodated without changing the \emph{angular} scale of the sound horizon,
\begin{equation}
\label{eq:thetas}
    \theta_s = \frac{r_s (z_*)}{D_A(z_*)},
\end{equation}
which is measured to $0.03\%$ precision by the \emph{Planck} 2018 CMB data \cite{Aghanim:2018eyx}. 

Particle physics realizations of this scenario are strongly constrained by simple considerations of Eqs.~\eqref{eq:rs}, \eqref{eq:DA}, and \eqref{eq:thetas}. The sound horizon at last scattering in Eq.~\eqref{eq:rs} is dominated by contributions near the lower bound of the integral, and thus is  primarily sensitive to the evolution of $H(z)$ at times shortly before recombination. In addition, the magnitude of the Hubble tension ($\approx 10\%$) implies via Eqs.~\eqref{eq:DA} and~\eqref{eq:thetas} that the increase in the expansion rate just prior to recombination must also be of order $10\%$. Translated into natural units, this implies an energy density $\sim {\rm eV}$, three orders of magnitude larger than the present-day vacuum energy density, must be present when the Hubble parameter was $H \sim 10^{-28}$ eV. The final piece of the EDE scenario is that this extra energy density must rapidly decay after last-scattering, so as not to directly impact the formation of structure at late times.

The requisite dynamics can be straightforwardly realized in scalar field models. From the equation of motion of a canonical massive scalar field in an expanding universe,
\be
\ddot{\phi} + 3 H \dot{\phi} + m^2 \phi =0,
\ee
one may appreciate two distinct regimes of dynamics: if $m\ll H$, the above has an approximate solution $\phi = \phi_i$ with $\phi_i$ a constant. The energy density of the scalar field is dominated by the potential energy, $V = m^2 \phi_i^2/2$, and hence gravitates as dark energy. In the opposite limit, $m \gg H$, the field undergoes damped oscillations, $\phi \simeq \phi_i a^{-3/2} \cos m t$, and the energy density redshifts like matter, $\rho \propto a^{-3}$.

In a $\Lambda$CDM universe, the boundary of these asymptotic regimes, $m \simeq H$, sets the time of decay of the `early dark energy' stored in the $\phi$ field. In order to play a role in addressing the Hubble tension, the requisite timing of the decay thus demands that the scalar field be extremely light, $m \sim 10^{-28} {\rm eV}$. In the context of particle physics, the only model construction of such a light field is the axion \cite{Peccei:1977hh,Wilczek:1977pj,Weinberg:1977ma}, which obtains a mass only through a periodic potential $V \sim m^2 f^2 \cos \phi/f$ generated by non-perturbative effects.

On the other hand, the EDE field must decay as fast or faster than radiation, while in the simple example above, its energy density redshifts as matter. To simultaneously achieve this aspect, one may generalize the model to include multiple fields or new decay channels. For example, the EDE may be converted into kinetic energy \cite{Alexander:2019rsc,Lin:2019qug}, gravitational waves \cite{Niedermann:2019olb}, relativistic particles \cite{Kaloper:2019lpl}, or gauge fields \cite{Berghaus:2019cls}, among other possibilities.

\begin{figure}
\begin{center}
\includegraphics[width=\linewidth]{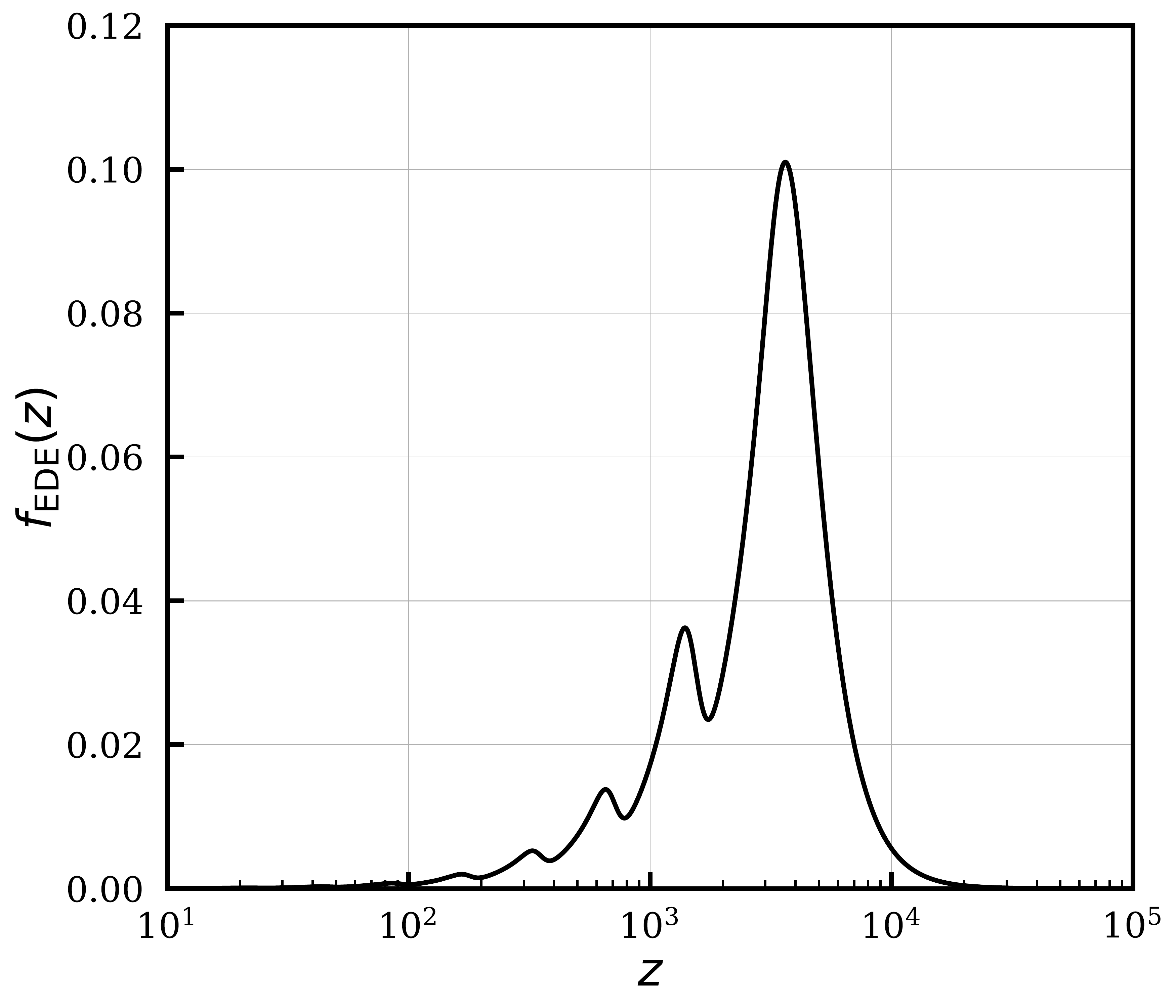}
\caption{Fraction of the cosmic energy density in the EDE field as a function of redshift, for the parameters in Eq.~\eqref{hillparams}.}
\label{fig:EDEn3}
\end{center}
\end{figure} 

A well-studied possibility is to generalize the axion potential to include higher-order harmonics. The proposal of \cite{Poulin:2018cxd} is to consider a single scalar field with potential (see also~\cite{Kamionkowski:2014zda}),
\be
V = V_0 \left( 1 - \cos (\phi/f)\right)^n \;\; , \;\; V_0 \equiv m^2 f^2 \, .
\label{eq.PoulinEDE}
\ee
The parameter $n$ serves to set the decay rate of the EDE. The minimum of the potential in Eq.~\eqref{eq.PoulinEDE} is locally $V\sim \phi^{2n}$, and the energy density of oscillations in this minimum has an equation of state~\cite{PhysRevD.28.1243},
\be
w_{\phi} = \frac{n-1}{n+1} .
\ee
For $n=2$, the initial energy stored in the field (i.e.,~the EDE) redshifts as radiation ($\propto a^{-4}$), and for $n\rightarrow \infty$ it redshifts as kinetic energy ($\propto a^{-6}$). Recent results indicate that $n=3$ provides the best-fit integer value to cosmological data, although the preference is fairly weak~\cite{Smith:2019ihp}.

The cosmological dynamics relevant to the Hubble tension can be succinctly described by two effective parameters, $z_c$ and $f_{\rm EDE}$, corresponding to the redshift $z_c$ at which the EDE makes its peak contribution $f_{\rm EDE}$ to the total energy density of the universe. Along with an initial condition $\theta_i \equiv \phi_i/f$ and the exponent $n$, these parameters determine the timing, relative amount, and decay rate of the EDE component.

As a fiducial example, we consider the best-fit parameters found in \cite{Hill:2020osr} in the fit of the $n=3$ EDE model in Eq.~\eqref{eq.PoulinEDE} to the CMB power spectra, CMB lensing, BAO, RSD, Type Ia supernovae, and the SH0ES $H_0$ measurement. 
The parameters are
\begin{align}
    \label{hillparams}
     H_0&=71.15 \, {\rm km/s/Mpc} ,&100\omega_b=2.286  \\
     \omega_{\rm cdm}&=0.12999 ,& {\rm ln}\,  10^{10} A_s = 3.058, \nonumber \\  n_s&=0.9847 ,&\tau_{\rm reio}=0.0511 \nonumber \\
      f_{\rm EDE}&=0.105 ~~~~~ \log_{10}(z_c)=3.59  &\theta_i = 2.71\,, \nonumber\\
     \Omega_m &=0.303 
      ~~~~~ \sigma_8 = 0.8322
     &S_8  = 0.8366,
      \nonumber \\ \nonumber
   f\sigma_8|_{z=0.38}  &=0.482  &f\sigma_8|_{z=0.61} =0.477\,.
\end{align}
The cosmological evolution of the energy density in the EDE field, i.e., $f_{\rm EDE}(z)$, is shown in Fig.~\ref{fig:EDEn3}. At its peak, the EDE field comprises 10\% of the energy density of the universe. This is rapidly dissipated once the field begins to oscillate, and by $z=10^3$ it comprises less than 2\% of the energy density of the universe.

\begin{figure*}
    \centering
    \includegraphics[width=\linewidth]{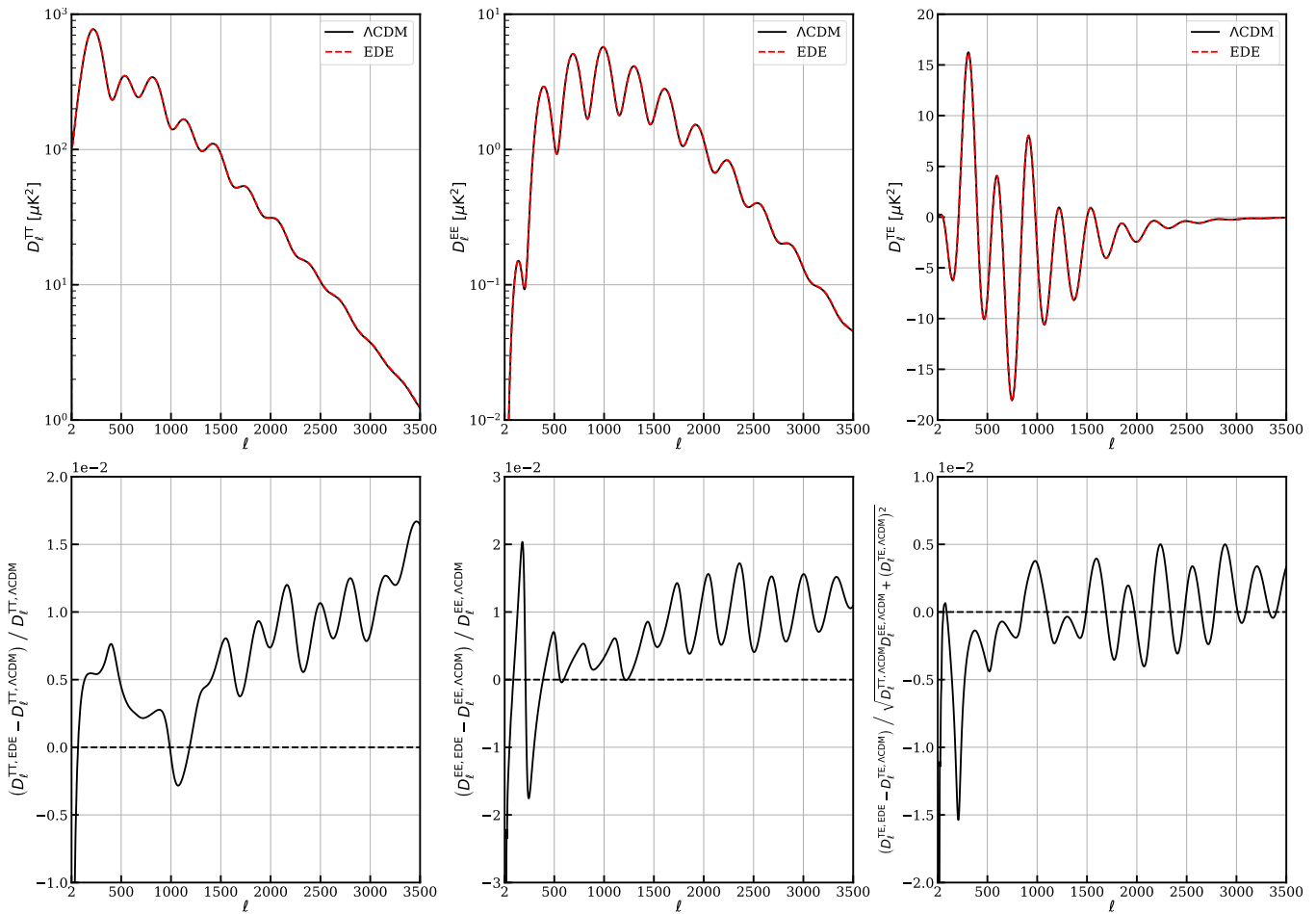}
    \caption{
    CMB TT (left panel), EE (middle panel), and TE (right panel) power spectra for $\Lambda$CDM (black, solid) and EDE (red, dashed), with $H_0 = 68.07$ km/s/Mpc and $H_0 = 71.15$ km/s/Mpc, respectively, and fractional difference between EDE and $\Lambda$CDM (bottom). The fractional difference for TT and EE is normalized to the $\Lambda$CDM spectra, while TE has been normalized by the variance to accommodate the zero crossings in this spectrum. The model parameters are given in Eqs.~\eqref{hillparams} and~\eqref{hillparamsLCDM} for EDE and $\Lambda$CDM, respectively, corresponding to the best-fit parameters from \cite{Hill:2020osr} in the fit to primary CMB, CMB lensing, BAO, RSD, SNIa, and SH0ES data.
   }
    \label{fig:CMB_TT}
\end{figure*}

As promised, this EDE model preserves the fit to the CMB power spectra to a remarkable degree.  The proper inclusion of the EDE perturbations in the calculation is crucial for achieving this result~\cite{Poulin:2018cxd}. As a basis for comparison, we consider $\Lambda$CDM fit to the same combination of data sets, with best-fit parameters given by \cite{Hill:2020osr}
\begin{align}
     \label{hillparamsLCDM}
     H_0&=68.07 \, {\rm km/s/Mpc} &   100\omega_b&=2.249, \\
     \omega_{\rm cdm}&=0.11855 & {\rm ln} \,10^{10} A_s &= 3.047, \nonumber \\  n_s&=0.9686 &    \tau_{\rm reio}&=0.0566, \nonumber\\
     \Omega_m &=0.306 
      ~~~~~~~~\sigma_8 = 0.808
      & S_8  &=0.816,
      \nonumber \\
       \nonumber
      f\sigma_8|_{z=0.38}  &=0.47  &f\sigma_8|_{z=0.61} &=0.464\,.
\end{align}
The CMB TT, EE, and TE power spectra in $\Lambda$CDM with parameters from Eq.~\eqref{hillparamsLCDM} and EDE with parameters from Eq.~\eqref{hillparams} are shown in Fig.~\ref{fig:CMB_TT}. The difference is not visible by eye, despite the differing $H_0$ values, $H_0=68.07 \, {\rm km/s/Mpc}$ and $H_0=71.15 \, {\rm km/s/Mpc}$ for $\Lambda$CDM and EDE, respectively.

\section{EDE meets LSS}
\label{sec:EDExLSS}

As we have observed, the EDE extension of $\Lambda$CDM can accommodate a larger $H_0$ while maintaining an excellent fit to the primary CMB anisotropies. However, this is achieved through a substantial shift in the standard $\Lambda$CDM parameters when fitting the EDE model, as can be appreciated by comparing Eqs.~\eqref{hillparams} and \eqref{hillparamsLCDM}. As discussed in detail in \cite{Hill:2020osr}, these parameter shifts leave an imprint on cosmological observables beyond the CMB primary anisotropies and $H_0$.

The implications for LSS observations can be understood already within linear perturbation theory. For example, the relative increase in the physical dark matter density leads to an increase in the $\sigma_8$ parameter, the RMS linear-theory mass fluctuation in a sphere of radius $8 \, {\rm Mpc}/h$ at $z=0$,
\begin{equation}
    \left(\sigma_8 \right)^2  \equiv \int {\rm d}\log k \, \frac{k^3}{2\pi^2} P_{\rm lin}(k) \, W^2(k R).
\end{equation}
This increase in $\sigma_8$ in turn leads to a relative increase in the related $S_8$ parameter, $S_8 \equiv \sigma_8 \left(\Omega_m/0.3 \right)^{0.5}$, worsening the known tension in $\Lambda$CDM  parameter inferences between CMB and LSS observations~(e.g.,~\cite{Abbott:2017wau,Hikage:2018qbn,2020A&A...633A..69H}).

Similarly, the combination $f\sigma_8(z)$, where $f$ is the logarithmic growth rate, defined as the logarithmic derivative of the linear growth factor $D(a)$,
\begin{equation}
    f = \frac{{\rm d} \ln D}{{\rm d} \ln a} ,
\end{equation}
exhibits an increase at the 2-3\% level across a range of redshift \cite{Hill:2020osr}. The growth rate determines the linear-theory prediction for the divergence of velocity perturbations, $\theta=- f \delta_m$, the power spectrum of which contributes to the anisotropic galaxy clustering power spectrum,  typically encoded in the parameter combination $f\sigma_8$. This is probed observationally through the measurement of RSD~\cite{1987MNRAS.2271K}.

Finally, the shifts in the standard cosmological parameters needed to fit the CMB data in the EDE scenario can be used themselves to constrain this model via their signature in LSS observables. For example, the physical cold dark matter density $\omega_{\rm cdm}$ increases by over 5\% when going from $\Lambda$CDM to EDE best-fit parameters (see Eqs.~\eqref{hillparams} and~\eqref{hillparamsLCDM}).  This increase is needed to compensate for the early-ISW-induced growth suppression caused by the EDE.  On the other hand, as shown in~\cite{DAmico:2019fhj,Ivanov:2019pdj}, $\omega_{\rm cdm}$ can be measured directly from the shape of the galaxy power spectrum, without any information from the CMB. As we will see, this shape information and constraints on $\omega_{\rm cdm}$ from the BOSS data will play an important role in constraining the EDE model.

\begin{figure*}[t]
    \centering
    \includegraphics[width=0.49\textwidth]{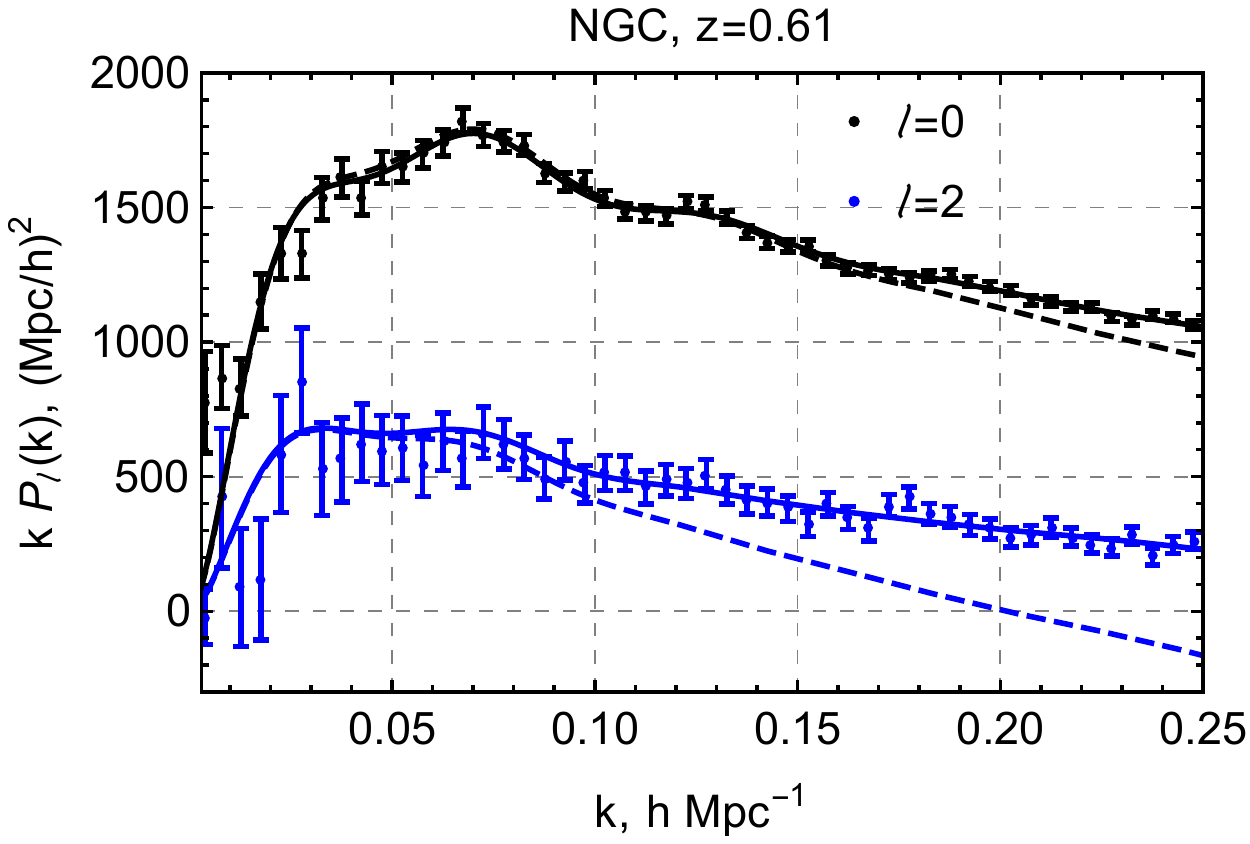}
    \includegraphics[width=0.49\textwidth]{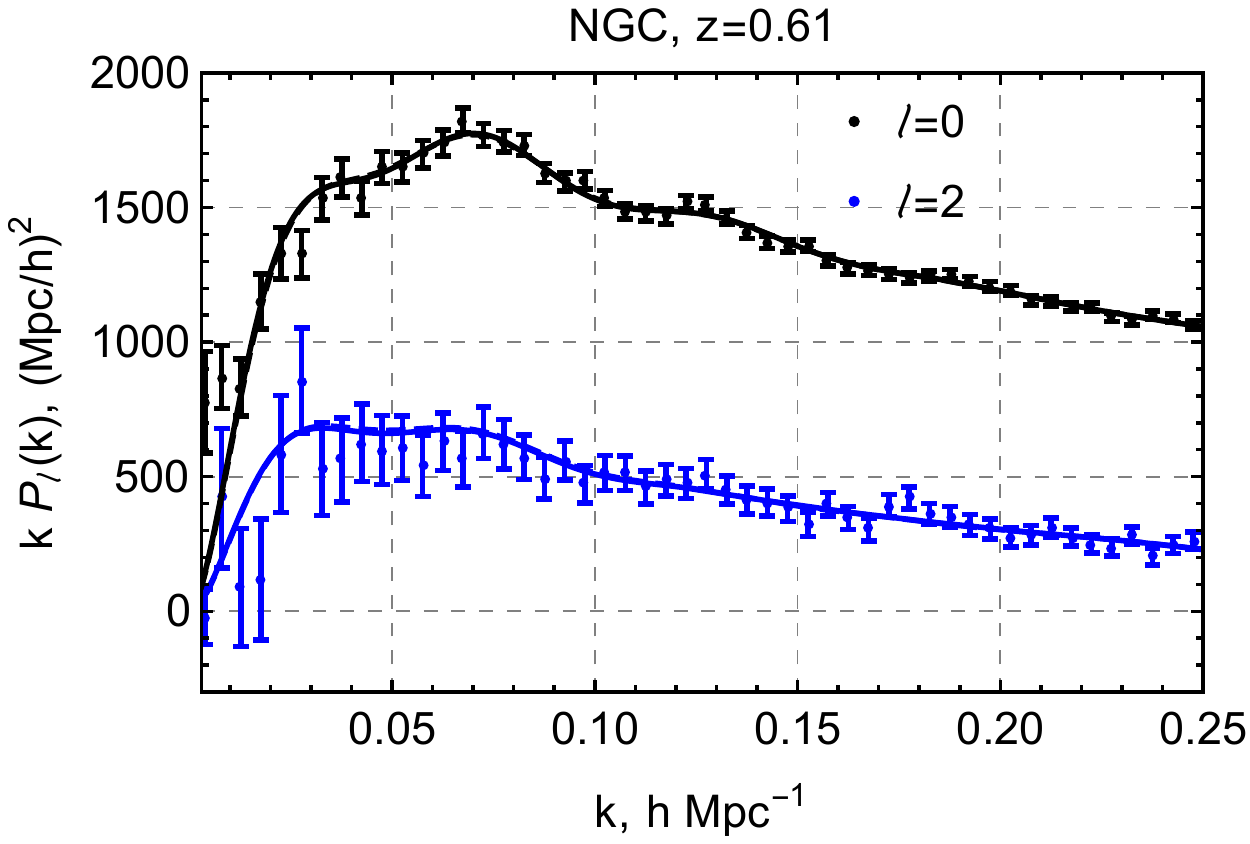}
        \caption{
        Multipoles of the galaxy power spectra at $z=0.61$, before (left panel) and after
    (right panel) marginalizing over nuisance parameters, along with the high-$z$ NGC BOSS data.
    The predictions of the $\Lambda$CDM model are shown with solid curves, while the
the EDE predictions are shown with dashed curves. In the right panel (after marginalizing over nuisance parameters) the curves cannot be distinguished by eye; the fractional difference in these curves is shown in Fig.~\ref{fig:multipoles}.     }
    \label{fig:spectra}
\end{figure*}

\begin{figure*}
    \centering
        \includegraphics[width=0.49\textwidth]{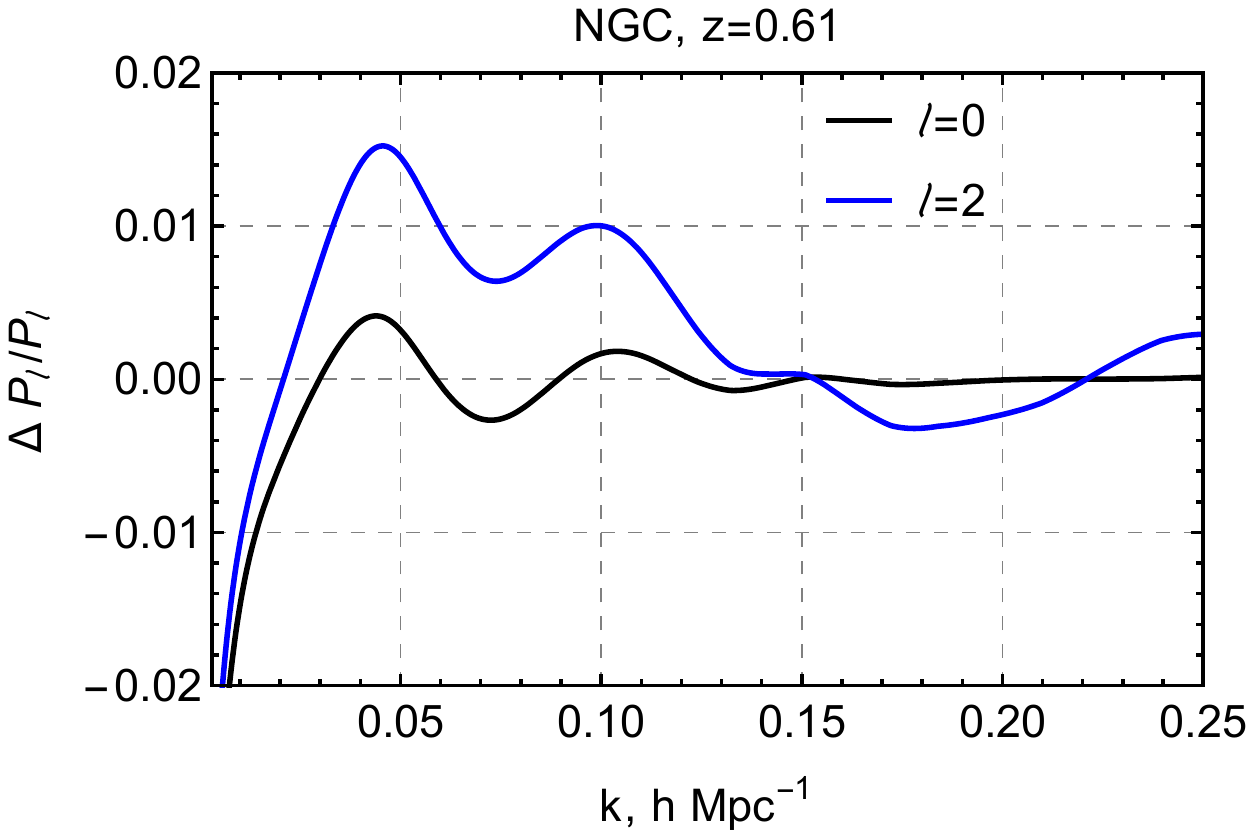}
    \includegraphics[width=0.49\textwidth]{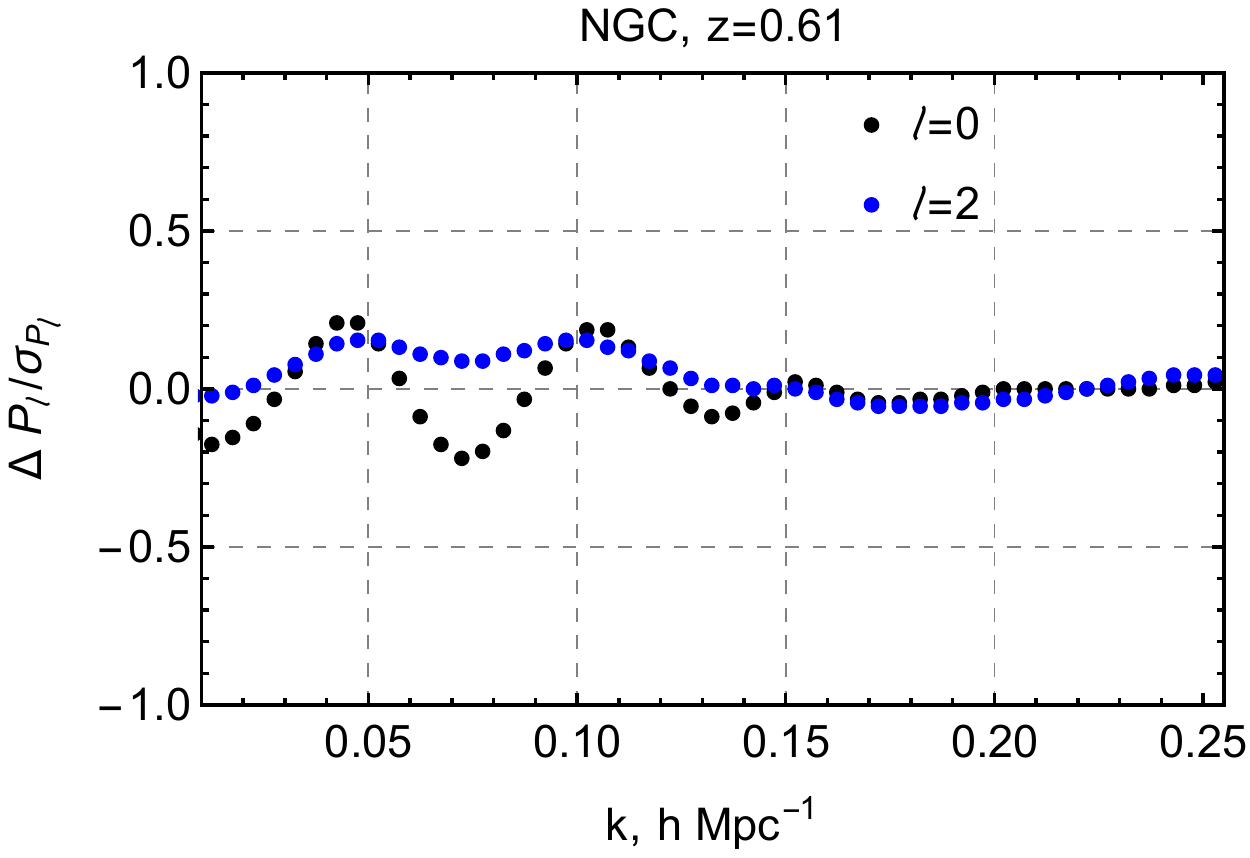}
    \caption{Multipoles of the galaxy power spectrum at $z=0.61$, after marginalizing over nuisance parameters as in the right panel of Fig.~\ref{fig:spectra}. {\it Left panel}: Fractional difference between $\Lambda$CDM and EDE: $\Delta P/P\equiv(P^{\rm EDE}- P^{\Lambda{\rm CDM}})/P^{\Lambda{\rm CDM}}$.  The monopole features a $0.3\%$ pattern produced by the mismatch in the shape of the BAO wiggles between the two models, whereas the quadrupole exhibits a $\mathcal{O}(2\%)$ fractional difference at low $k$. {\it Right panel}: Fractional difference in units of the BOSS data error bar for every wavenumber bin: $\Delta P/\sigma_P$. (Note that the neighboring $k$ bins are correlated). The biggest discrepancy is observed in the shape and position of the BAO wiggles in the monopole; see the main text for details.}
 \label{fig:multipoles}
\end{figure*}

While linear perturbation theory is very useful to gain some insight into the effects of the EDE model on the large-scale clustering, in order to make full use of the breadth of LSS observations it is required to go beyond linear theory, and instead make predictions for the power spectrum in the {\it non-linear} regime. This regime is complicated by various factors, for example, non-linear evolution of the matter density field, the relation between this field and its luminous tracers such as galaxies, and baryonic effects. There are two common approaches to deal with non-linearities: numerical $N$-body (or fully hydrodynamical) simulations and perturbation theory. 

The advantage of $N$-body simulations is their ability, in principle, to describe all scales including highly non-linear ones. However, achieving high-precision $N$-body simulations  requires significant computational resources, and these simulations are nonetheless affected by uncertainties in galaxy formation models on small scales. A computational compromise is to use $N$-body simulations to calibrate fitting formulas for dark matter non-linearities, such as HALOFIT \cite{Smith:2002dz,Takahashi2012}, and to use phenomenological models for the galaxy bias. This approach is standard in the analyses of photometric survey data, which are affected by non-linear effects down to small scales. However, this too has its limitations: since the fitting formulas are not derived from first principles, their applicability to extensions of $\Lambda$CDM is, at best, questionable. In the context of EDE, the use of HALOFIT has been justified \cite{Hill:2020osr} by noting the smallness of the deviation from a $\Lambda$CDM cosmology in the allowed region of EDE parameter space, and by a close match of posterior distributions in the fit to the DES data \cite{Abbott:2017wau}, namely those obtained with the non-linear matter power spectrum computed with HALOFIT, and those obtained when replacing the DES data with a prior on $S_8$ (which by definition requires only linear theory). This comparison suggests that HALOFIT is precise enough for the photometric LSS data sets available today. 

Another approach to galaxy clustering is nonlinear perturbation theory, whose formulation was recently finalized in the framework of the Effective Field Theory of Large-Scale Structure \cite{Baumann:2010tm,Carrasco:2012cv}. Even though the EFT is intrinsically limited to a narrow range of wavenumber $k\lesssim 0.5~h$/Mpc, it gives unprecedented accuracy in the description of non-linear clustering within these scales.  Perturbative techniques are typically used in spectroscopic surveys, which aim at reconstructing the three-dimensional matter distribution, allowing for a clear separation of scales. Importantly, the EFT can robustly account for all effects shaping the observed map of LSS: non-linearities in the underlying matter field, galaxy bias, RSD, fingers-of-God, baryonic feedback, etc. The EFT provides us with a general large-scale description of a cosmological fluid that possesses rotation symmetry and obeys the equivalence principle. Both these symmetries are present in the EDE scenario, which makes the EFT an adequate tool to analyze the BOSS redshift-space clustering data in this model. In the rest of this paper we will use the one-loop EFT model parametrized as in~\cite{Ivanov:2019pdj} (see also~\cite{DAmico:2019fhj,Chudaykin:2020aoj} and references therein for more detail about theoretical modelling and the EFT approach).\footnote{This model requires 7 nuisance parameters for each BOSS data chunk. These are galaxy bias parameters $b_1,b_2,b_{\mathcal{G}_2}$, the constant shot-noise contribution $P_{\rm shot}$, the quadratic counterterms $c_0^2,c_2^2$, and the higher-order fingers-of-God counterterm $\tilde{c}$. This parameter also accounts for the effect of fiber collisions on the measured quadrupole moment.}

It is instructive to look at the effects of the EDE on the observed redshift-space galaxy power spectrum. We will consider its decomposition into multipole moments, in particular, the $\ell=0$ isotropic component (monopole) and the leading contribution to anisotropic galaxy clustering, the $\ell=2$ moment (quadrupole). These are the spectra that will be used in our data analysis. Let us focus on the power spectrum at $z=0.61$, which corresponds to the high-$z$ NGC BOSS data.  As a basis for comparison, we consider the model parameters given in Eqs.~\eqref{hillparams} and~\eqref{hillparamsLCDM} for EDE and $\Lambda$CDM, respectively, corresponding to the best-fit parameters from~\cite{Hill:2020osr} in the fit to primary CMB, CMB lensing, BAO, RSD, SNIa, and SH0ES data. 

If all nuisance parameters in the galaxy power spectrum are fixed to the same values both in the EDE and in the $\Lambda$CDM model, there are noticeable differences between the two, which can be seen in the left panel of Fig.~\ref{fig:spectra}. There we show the theoretical spectra for $\Lambda$CDM, whose nuisance parameters were fit to the data, along with the EDE predictions evaluated for the same set of nuisance parameters. However, most of this difference can be absorbed into the nuisance parameters, when they are allowed to vary. This effect is illustrated in the right panel of Fig.~\ref{fig:spectra}, which shows the same spectra after fitting the nuisance parameters for each cosmology separately. Note that the difference between these parameters needed to  compensate the mismatch between $\Lambda$CDM and EDE  is $\sim 10\%$, which is comparable to but smaller than the current precision with which these parameters are measured from the data or $N$-body simulations. 

The fractional difference between the two models (after absorbing the nuisance parameters) is shown in the left panel of Fig.~\ref{fig:multipoles}. One clearly sees that the monopole features a $0.3\%$ pattern produced by the mismatch in the shape and location of the BAO wiggles between the two models, whereas the quadrupole exhibits a $\mathcal{O}(2\%)$ fractional difference at low $k$'s. When normalized to the actual data error bars, the biggest discrepancy is observed in the shape of the BAO wiggles in the monopole, as seen in the right panel of Fig.~\ref{fig:multipoles}.  
The origin of this discrepancy can 
be understood as follows. 
The shape of the BAO wiggles is sensitive to $\omega_{\rm cdm}$ \cite{Ivanov:2019pdj}, which is different in the two models. Fig.~\ref{fig:multipoles}.  
This effect is more significant in the monopole because the non-linear
suppression of the BAO wiggles is weakest for this moment~\cite{Ivanov:2018gjr}, whilst the statistical error bars
are smallest. 
It is important to stress that the EDE model 
predicts $\sim 2\%$ larger velocity fluctuation 
amplitude $f\sigma_8$, 
which is constrained through
the correlation of the monopole and quadrupole moments.The excess 
in the quadrupole amplitude is clearly seen in the left panel of Fig.~\ref{fig:multipoles}.
It is worth mentioning that the oscillating residual observed in Fig.~\ref{fig:multipoles} is larger than the numerical inaccuracies
of the \texttt{CLASS-PT} code, 
which do not exceed the $0.1\%$ level~\cite{Chudaykin:2020aoj}.
It was shown in
Ref.~\cite{Nishimichi:2020tvu} 
that this level of accuracy is sufficient to ensure unbiased cosmological constraints even for a galaxy survey many times larger than BOSS.

All in all, the difference
in the $\chi^2-$statistic
between the two models for the high-$z$ NGC sample is $\Delta \chi^2 = 2.1$. Repeating the same exercise for the three other BOSS data chunks, we find the cumulative $\Delta \chi^2 = 2.5$. This simple comparison suggests that the BOSS data can improve the constraints on the EDE model in combination with other datasets, such as the CMB.

Note that this comparison is based on the best-fit
parameters from Eq.~\eqref{hillparams} obtained without the full-shape BOSS data.
This set of parameters may no longer be a best-fit after the addition of this data, which is expected to break some
parameter degeneracies.
What is relevant to estimate 
the evidence for 
EDE as a valid resolution
to the Hubble tension
is the difference 
between the actual best-fits 
at $f_{\rm EDE}\approx 0$
and $f_{\rm EDE}\approx 0.1$,
obtained in the presence of the BOSS data. This will be addressed
in the following sections, and in particular, in Appendix~\ref{app:like}.
The main purpose of our exercise
presented in this section
is to
illustrate, at the qualitative level,
that the BOSS data has the potential to improve the EDE constraints.

\section{Constraints on the EDE Scenario}
\label{sec:constraints}

In the following we use a combined Einstein-Boltzmann code comprised of \texttt{CLASS\_EDE}~\cite{Hill:2020osr} and \texttt{CLASS-PT}~\cite{Chudaykin:2020aoj} (both extensions of \texttt{CLASS}~\cite{2011JCAP...07..034B}), interfaced with the Monte Carlo sampling code \texttt{Monte Python} \cite{Brinckmann:2018cvx,Audren:2012wb}. Each of these codes is publicly available. We perform Markov chain Monte Carlo (MCMC) analyses, sampling from the posterior distributions using the Metropolis-Hastings algorithm \cite{LewisBridle2002,Lewis2013,Neal2005}, with a Gelman-Rubin \cite{Gelman:1992zz} convergence criterion $R-1 < 0.15$ (unless otherwise stated).\footnote{Our convergence criterion is somewhat weaker than the typically used criterion $R-1<0.1$. We have chosen to use $R-1<0.15$ as a compromise because the parameter exploration of the joint BOSS+Planck likelihood turned out to be computationally expensive. The quantile criterion applied to our sample of $20$ million accepted Monte-Carlo steps shows that the relative variance between the parameter errors from different sub-samples is less than $10\%$.}  We analyze the MCMC chains using both \texttt{MontePython} and
\texttt{GetDist}~\cite{GetDist},\footnote{\url{https://github.com/cmbant/getdist}} which give very similar results.

It is important to note that there are two commonly used ways to present $68\%$ marginalized confidence intervals in the case of two-tailed limits. The first is to display a limit such that $32\%$ of samples are outside the limit range, i.e., that either tail contains $16\%$ of the samples. The second option is to quote an interval between two points with highest equal marginalized probability density (called the ``credible interval''). The two approaches yield identical confidence intervals for the Gaussian distribution, but can notably differ if the distribution is skewed, which is the case for the parameter posterior distributions of the EDE model. Given this uncertainty, we use the second approach in the main part of the paper and present the alternative estimates obtained with the equal-tail method in Appendix~\ref{app:full}. Note that the equal-tail method was used in~\cite{Hill:2020osr}. In either case, the limit will be presented here as 
\[
\text{mean}^{+\text{(upper $68\%$ limit - mean)}}_{-\text{(mean - lower $68\%$ limit)}}\,.
\]
As far as $f_{\rm EDE}$ is concerned, we will see that its posterior is peaked at the
lower boundary $f_{\rm EDE}=0.001$,
and hence we will
quote the 95$\%$ CL upper limit, i.e., the point where the cumulative probability distribution function equals $0.95$.

We impose uniform priors on the EDE parameters: $f_{\rm EDE} = [0.001, 0.5]$, $\log_{10}(z_c) = [  3., 4.3]$ and $\theta_{i}= [  0.1, 3.1]$. For a detailed discussion of priors in the context of EDE, see \cite{Hill:2020osr}.  We fix $n=3$ throughout, as the data only weakly constrain this parameter~\cite{Smith:2019ihp}.  We assume broad uniform priors on the standard $\Lambda$CDM parameters, and, following the \emph{Planck} convention, we fix the sum of the neutrino masses to be 0.06 eV, assuming one massive eigenstate and two massless eigenstates. We fix the effective number of relativistic species  $N_{\rm eff} = 3.046$.   

Constraints on EDE from the primary CMB anisotropies alone were first reported in \cite{Hill:2020osr}, which we take as the starting point for our analysis.  That work found no evidence for EDE in the {\it Planck} 2018 primary CMB temperature and polarization data, obtaining a 95\% CL upper bound $f_{\rm EDE} < 0.087$, below typical values needed to fully resolve the Hubble tension (see, e.g., Fig.~\ref{fig:CMB_TT} or \cite{Poulin:2018cxd,Smith:2019ihp}, which indicate that $f_{\rm EDE} \simeq 0.10$-$0.12$ could resolve the Hubble tension). Ref.~\cite{Hill:2020osr} found $H_0 = 68.29 ^{+1.02} _{-1.00}$ km/s/Mpc in the EDE fit to \emph{Planck} alone, slightly larger and with a considerably larger error bar than the corresponding $\Lambda$CDM value, $H_0 = 67.29 \pm 0.59$ km/s/Mpc. The $S_8$ parameter was found to be $S_8=0.839 \pm 0.017$, again slightly larger than the $\Lambda$CDM value, $S_8=0.833\pm 0.016$. For a complete discussion and plots of the posterior distributions, we refer the reader to \cite{Hill:2020osr}. When plotting {\it Planck}-only results in the figures below, we use the MCMC chains from \cite{Hill:2020osr}.

\subsection{Datasets}
\label{sec:data}

For the CMB, we use the final {\it Planck} 2018 TT+TE+EE+low $\ell$+lensing 
likelihood \cite{Planck2018likelihood}. We follow the standard analysis routine for this likelihood and vary all necessary nuisance parameters required to account for 
observational and instrumental uncertainties.

For BOSS, we use the data from final release DR12~\cite{Alam:2016hwk}, implemented as a joint full-shape+BAO likelihood in Ref.~\cite{Philcox:2020vvt}. We refer the reader to Refs.~\cite{Ivanov:2019pdj,Chudaykin:2020aoj} for details of the pre-reconstruction full-shape galaxy power spectrum likelihood (based on the EFT described earlier) and to Ref.~\cite{Philcox:2020vvt} for details of the post-reconstruction BAO extraction and the BAO-FS covariance matrix. The likelihood includes pre- and post-reconstruction anisotropic galaxy power spectrum multipoles  $\ell=0,2$ across two non-overlapping redshift bins with $z_{\rm eff}=0.38$ (low-$z$) and $z_{\rm eff}=0.61$ (high-$z$) observed in the North and South Galactic Caps (NGC and SGC, respectively). This yields four independent data chunks with a cumulative volume $\simeq 6~(h^{-1}{\rm Gpc})^3$. These data chunks have different selection functions, and hence require separate sets of nuisance parameters. We use wide conservative priors on the nuisance parameters, as in Ref.~\cite{Ivanov:2019pdj}.
We use the wavenumber range [$0.01,0.25$] $h$ Mpc$^{-1}$
for the pre-reconstruction
power spectra in the FS part of the likelihood and 
[$0.01,0.3$] $h$ Mpc$^{-1}$
for the BAO measurements from the post-reconstruction spectra.

Finally, we include additional LSS data from photometric surveys in the analysis. In particular, we consider the DES photometric galaxy clustering, galaxy-galaxy lensing, and cosmic shear measurements~\cite{Abbott:2017wau}, in addition to weak gravitational lensing measurements from KV-450~\cite{Hildebrandt:2016iqg,2020A&A...633A..69H} and HSC~\cite{Hikage:2018qbn}. It was demonstrated in \cite{Hill:2020osr} that the DES-Y1 data set, namely the ``3x2pt'' likelihood from two-point correlations of photometric galaxy clustering, galaxy-galaxy lensing, and cosmic shear, is well approximated in the EDE analysis (with {\it Planck} and other data sets) by a Gaussian prior on $S_8$ corresponding to the DES measurement. Guided by this, we include DES-Y1, as well as KV-450 and HSC, via priors on $S_8$.  For DES we use the result $S_8 = 0.773 ^{+0.026} _{-0.020}$; for KV-450, we use the result from \cite{2020A&A...633A..69H}: $S_8 = 0.737^{+0.040}_{-0.036}$; and for HSC, we use the result from \cite{Hikage:2018qbn}: $S_8 = 0.780^{+0.030}_{-0.033}$. The inverse-variance weighted combination of these measurements gives $S_8 = 0.770 \pm 0.017$. In what follows, we will refer to this combination simply as ``$S_8$.''

\begin{table*}[htb!]
Constraints from \emph{Planck} 2018 data  + BOSS DR12 \vspace{2pt} \\
  \centering
  \begin{tabular}{|l|c|c|}
    \hline\hline Parameter &$\Lambda$CDM~~&~~~EDE ($n=3$) ~~~\\ \hline \hline

    {\boldmath$\ln(10^{10} A_\mathrm{s})$} & $3.043 \, (3.034) \, \pm 0.014$ & $3.047 \, (3.049) \, \pm 0.014$ \\

    {\boldmath$n_\mathrm{s}$} & $0.9656 \, (0.9655) \, \pm 0.0037 $ & $0.9696 \, (0.9717)^{+0.0046}_{-0.0068}$ \\

    {\boldmath$100\theta_\mathrm{s}$} & $1.04185 \, (1.04200) \, \pm 0.00029 $ & $1.04172 \, (1.04126) \, \pm 0.00032$\\

    {\boldmath$\Omega_\mathrm{b} h^2$} & $0.02241 \, (0.02233) \, \pm 0.00014 $ &  $0.02255 \, (0.02245) \, \pm 0.00018$ \\

    {\boldmath$\Omega_\mathrm{cdm} h^2$} & $ 0.1192\, (0.1191)_{-0.00095}^{+0.00087}$ & $0.1215 \, (0.1243)^{+0.0013}_{-0.0029}$ \\

    {\boldmath$\tau_\mathrm{reio}$} & $0.0546 \, (0.0503)_{-0.0072}^{+0.0065} $ & $0.0553 \, (0.0543)_{-0.0075}^{+0.0069}$\\

    {\boldmath$\mathrm{log}_{10}(z_c)$} & $-$ & $3.71 \, (3.52)^{+0.26}_{-0.33}$ \\

    {\boldmath$f_\mathrm{EDE} $} & $-$ & $< 0.072 \, (0.047)$\\

    {\boldmath$\theta_i$} & $-$ & $2.023(2.734)_{-0.34}^{+1.1} $\\

    \hline

    $H_0 \, [\mathrm{km/s/Mpc}]$ & $67.70 \, (67.56) \, \pm 0.42$ & $68.54 \, (68.83)^{+0.52}_{-0.95}$ \\

    $\Omega_\mathrm{m}$ & $0.3105 \, (0.3112)_{-0.0058}^{+0.0053}$ & $0.3082 \,(0.3120)_{-0.0057}^{+0.0056}$ \\

    $\sigma_8$ & $0.8077 \, (0.8039) \, \pm 0.0058$ & $0.8127 \, (0.8195)_{-0.0091}^{+0.0072}$ \\
    
    $S_8$ & 
    $0.822 \, (0.819) \, \pm 0.010$
    & 
    $0.824 \, (0.827)\, \pm 0.011$
    \\
    \hline
  \end{tabular} 
  \caption{The mean (best-fit) $\pm1\sigma$ constraints on the cosmological parameters in $\Lambda$CDM and in the EDE scenario with $n=3$, as inferred from the combination of BOSS FS+BAO and \emph{Planck} 2018 TT+TE+EE+low $\ell$+lensing data. The upper limit on $f_{\rm EDE}$ is quoted at 95\% CL. The EDE component is not detected here; a $68\%$ confidence limit yields $f_\mathrm{EDE} = 0.025_{-0.025}^{+0.0061}$, i.e., consistent with zero. The EDE value of $H_0$ is in $3.6\sigma$ tension with the SH0ES measurement  ($H_0=74.03\pm1.42$ km/s/Mpc).}
  \label{table:params-P18+BOSS}
\end{table*}

\subsection{EDE Meets BOSS: Constraints on EDE from the CMB and BOSS FS+BAO}
\label{sec:BossFS}

We perform a joint analysis of the CMB temperature, polarization, and lensing data combined with the BOSS DR12 full-shape and BAO likelihood. The non-linear spectra are computed using \texttt{CLASS-PT}  \cite{Chudaykin:2020aoj}, as discussed in Sec.~\ref{sec:EDExLSS}. Parameter constraints are given in Table \ref{table:params-P18+BOSS}, and the posterior distributions for a subset $(H_0,\omega_{cdm},\sigma_8,f_{\rm EDE},\log_{10}(z_c),\theta_i)$ are given in Fig.~\ref{fig:planck+BOSS}. The full triangle plot with all cosmological parameters can be found in Appendix~\ref{app:full}.

We find no evidence for EDE in this analysis, obtaining an upper bound $f_\mathrm{EDE} < 0.072$ at 95\% CL, representing a $\approx 20\%$ improvement on the constraint from the fit to CMB data alone \cite{Hill:2020osr}. This is accompanied by a downward shift in $S_8$ compared to that found from the CMB alone ($S_8=0.839 \pm 0.017$ \cite{Hill:2020osr}), and we find $S_8=0.822 \pm 0.010$. This is mirrored in $\Lambda$CDM, and we find \mbox{$S_8=0.824\pm0.011$}, again smaller than that found in the fit to the CMB alone ($S_8=0.833\pm 0.016$). This is driven by parallel shifts in $\sigma_8$ and $\Omega_m$; see Table~\ref{table:params-P18+BOSS}.

\begin{figure*}[ht!]
\begin{center}
\includegraphics[width=0.8\textwidth]{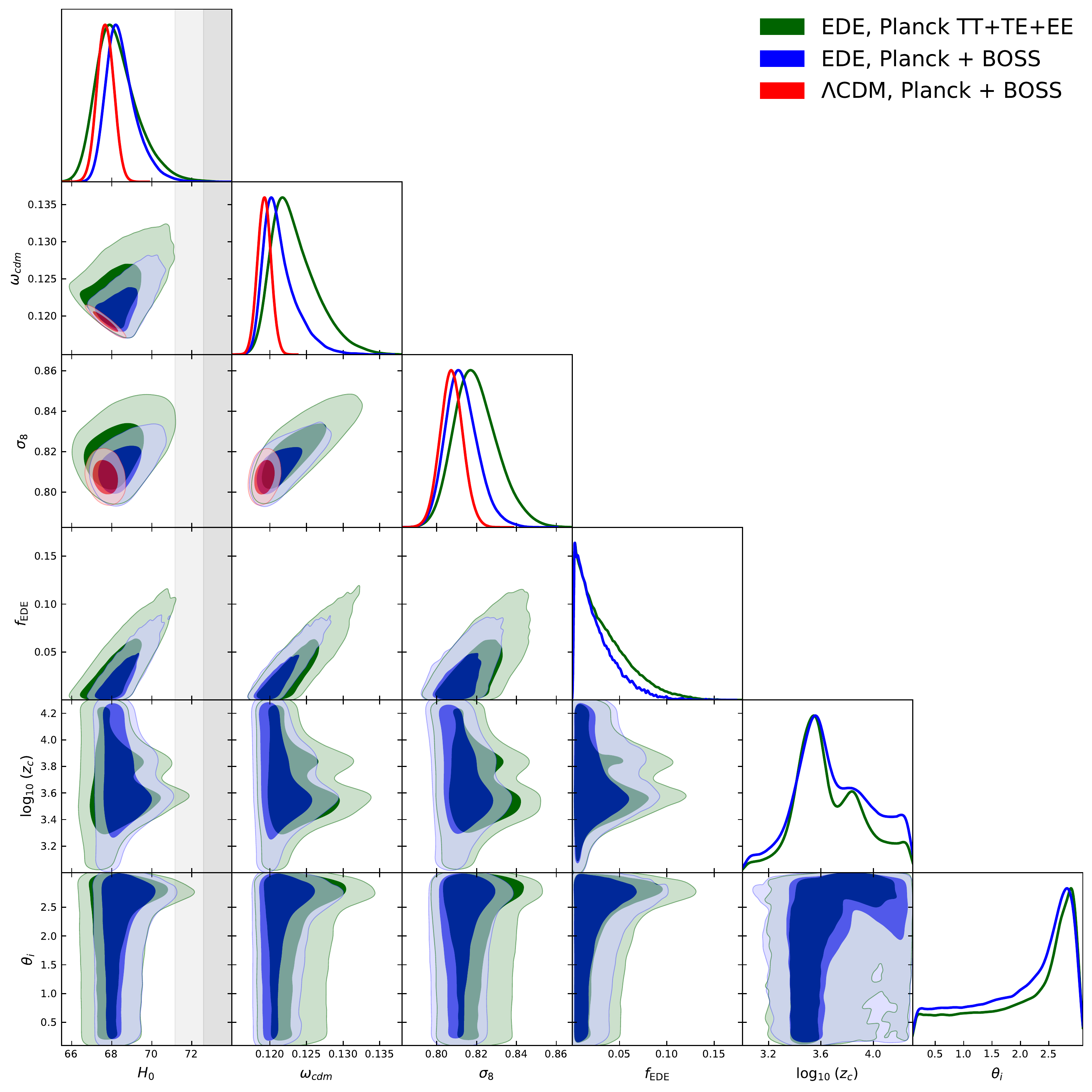}
\end{center}
\caption{Posterior distributions for the
parameters extracted from the joint {\it Planck} 2018 TT+TE+EE+low $\ell$+lensing + BOSS DR 12 (FS+BAO) likelihood.  For reference, we also display the  
 constraints from the Planck 2018 primary CMB data alone (TT+TE+EE), obtained in \cite{Hill:2020osr}. 
The dark-shaded and light-shaded contours mark $68\%$ and $95\%$ confidence regions, respectively. 
The gray band shows the $H_0$ measurement from SH0ES, for comparison ($1\sigma$ and $2\sigma$ regions in dark and light gray, respectively).\label{fig:planck+BOSS}}
\end{figure*}

We find that $H_0$ is shifted slightly upwards in both $\Lambda$CDM and EDE relative to its value in fits to the CMB alone. We find $H_0=68.54 ^{+0.52}_{-0.95}$ km/s/Mpc and $H_0=67.70 \pm 0.42$ in EDE and $\Lambda$CDM, respectively, both in significant tension with SH0ES (3.6$\sigma$ and 4.3$\sigma$, respectively). Both move towards the CMB-independent measurement $H_0 = 68.6\pm 1.1$km/s/Mpc from BOSS BAO data with a BBN prior imposed on $\omega_b$ \cite{Philcox:2020vvt}. This slight increase in $H_0$ when BAO is included is well known in the context of $\Lambda$CDM (see, e.g. Refs.~\cite{Alam:2016hwk,Aghanim:2018eyx}). This shift can be traced back to the fact that the effective volume-averaged distance\footnote{Defined as $D_V(z)\equiv (z D^2_A(z)H^{-1}(z))^{1/3}$, where  $D_A(z)$ and $H(z)$ are the comoving angular diameter distance and the Hubble parameter at redshift $z$, respectively.} $D_V$ imprinted in the BAO and the power spectrum turnover is a very weak function of the background cosmology and it probes, essentially, only $H_0$. That this is paralleled in EDE reflects not only the BAO preference for slightly larger $H_0$, but also the fact that EDE becomes indistinguishable from $\Lambda$CDM in the limit $f_{\rm EDE}\rightarrow0$. The constraints on the other standard $\Lambda$CDM parameters similarly track the $\Lambda$CDM constraints, e.g., the spectral index is reduced to $n_s=0.9696 ^{+0.0046} _{-0.0068}$, nearly identical to the value in $\Lambda$CDM fit to the same data set combination, $n_s=0.9656 \pm 0.0037$.

The timing of the EDE shows a preference for $\log_{10}(z_c)=3.71 ^{+0.26} _{-0.33}$, though there is substantial support on the boundary of the prior at $\log_{10}(z_c)=4.3$. Similarly, we find $\theta_i=2.023 ^{+1.1} _{-0.34}$,  with a posterior distribution that has substantial support at $\theta_i \simeq 0$.

Finally, it is interesting to compare the results of this subsection with those obtained from the combination of the {\it Planck} 2018 data and the standard BAO+RSD likelihood from BOSS. The details of this analysis are given in Appendix~\ref{app:fs8}.  The ``compressed'' BOSS likelihood does not appreciably narrow the {\it Planck}-only limits, and therefore does not confidently rule out the EDE as a resolution to the Hubble tension. We find $H_0 = 68.71_{-1.2}^{+0.69}$~km/s/Mpc, consistent with the upward shift in $H_0$ expected from the BOSS BAO likelihood.  However, due to the lack of shape information as compared to the EFT likelihood, the compressed likelihood allows for the increase of $\omega_{\rm cdm}$ associated with the upward shift of $H_0$ within the EDE model. This pushes the $f_{\rm EDE}$ posterior slightly away from the origin along the $H_0$-$f_{\rm EDE}$ degeneracy direction. The corresponding 95\% confidence limit $f_{\rm EDE}<0.096$ is compatible with the amount of the EDE required to account for the Hubble tension. This shows that the shape information beyond the commonly used $f\sigma_8$-BAO parametrization plays a crucial role in our {\it Planck}+BOSS constraints on the EDE scenario.

It is useful 
to check the extent to which our constraints
can be affected by the prior-volume effects. 
The standard way to assess prior volume effects is to compare the 1D marginalized  distribution for a given parameter 
with the 
average-likelihood profile of the samples for the same parameter~\cite{LewisBridle2002,Audren:2012wb,Aghanim:2018eyx}.
The latter is a smeared-out 
version of best-fit $\chi^2$ profile. 
If the mean likelihood is increased through the inclusion of additional model parameters, one may say that they indeed improve the fit to the data on average. 
This should be contrasted with 
the situation where 
the extra parameters need to be fine-tuned to obtain better fits.
The smearing is introduced exactly for this reason as it down-weights the $\chi^2$ values obtained as a result of fine-tuning.
This is an important test
as the EDE model has 3 extra parameters compared to the base $\Lambda$CDM model, which, at face value, should always improve the fit.
In our exercise, we have found that the mean likelihood profile for $f_{\rm EDE}$ is monotonically decreasing 
toward large $f_{\rm EDE}$, 
and its shape is close to the shape
of the 1D marginalized distribution; for more details, see Appendix~\ref{app:like}. This indicates that our limits are not driven by the prior volume effects.

\begin{table*}[htb!]
Constraints from \emph{Planck} 2018 data  + BOSS DR12  + $S_8$ from DES+KV-450+HSC\vspace{2pt} \\
  \centering
  \begin{tabular}{|l|c|c|}
    \hline\hline Parameter &$\Lambda$CDM~~&~~~EDE ($n=3$) ~~~\\ \hline \hline

    {\boldmath$\ln(10^{10} A_\mathrm{s})$} & $3.036 \, (3.039) \, \pm 0.014$ & $3.038 \, (3.034) \, \pm 0.014$ \\

    {\boldmath$n_\mathrm{s}$} & $0.9674 \, (0.9727) \, \pm 0.0037 $ & $0.9696 \, (0.9621)^{+0.0042}_{-0.0051}$ \\

    {\boldmath$100\theta_\mathrm{s}$} & $1.041945 \, (1.041966) \, \pm 0.00030 $ & $1.04178 \, (1.04176) \, \pm 0.00035$\\

    {\boldmath$\Omega_\mathrm{b} h^2$} & $0.02249 \, (0.02273) \, \pm 0.00013 $ &  $0.02259 \, (0.02243)^{+0.00016}_{-0.00018}$ \\

    {\boldmath$\Omega_\mathrm{cdm} h^2$} & $ 0.1182\, (0.1157)\pm 0.00081$ & $0.11958 \, (0.11951)^{+0.00096}_{-0.0018}$ \\

    {\boldmath$\tau_\mathrm{reio}$} & $0.0527 \, (0.0591)\pm 0.0067 $ & $0.0535 \, (0.0521)_{-0.0075}^{+0.0069}$\\

    {\boldmath$\mathrm{log}_{10}(z_c)$} & $-$ & $3.77 \, (4.24)^{+0.51}_{-0.33}$ \\

    {\boldmath$f_\mathrm{EDE} $} & $-$ & $< 0.0526 \, (0.0115)$\\

    {\boldmath$\theta_i$} & $-$ & $1.91(1.55)_{-0.47}^{+1.2} $\\

    \hline
    $H_0 \, [\mathrm{km/s/Mpc}]$ & $68.13 \, (69.28) \, \pm 0.38$ & $68.73 \, (67.92)^{+0.42}_{-0.69}$ \\

    $\Omega_\mathrm{m}$ & $0.3046 \, (0.2859)\pm 0.0049$ & $0.3024 \,(0.3091)\pm 0.0050$ \\

    $\sigma_8$ & $0.80204 \, (0.7947) \, \pm 0.0053$ & $0.8044 \, (0.8023)_{-0.0069}^{+0.0060}$ \\
    
    $S_8$ & 
    $0.8082 \, (0.7810) \, \pm 0.0086$
    & 
    $0.8075 \, (0.8143)\, \pm 0.0092$ 
    \\
    \hline
  \end{tabular} 
  \caption{
  The mean (best-fit) $\pm1\sigma$ constraints on the cosmological parameters in $\Lambda$CDM and in the EDE scenario with $n=3$, as inferred from the combination of BOSS FS+BAO, \emph{Planck} 2018 TT+TE+EE+low $\ell$+lensing, and DES+KV-450+HSC $S_8$ data.  The upper limit on $f_{\rm EDE}$ is quoted at 95\% CL. The EDE component is not detected here; a $68\%$ confidence limit is $f_\mathrm{EDE} = 0.019_{-0.019}^{+0.0040}$, i.e., consistent with zero. }
  \label{table:params-P18+BOSS-S8}
\end{table*}

\begin{figure*}[ht!]
\centering
 \includegraphics[width=0.8\textwidth]{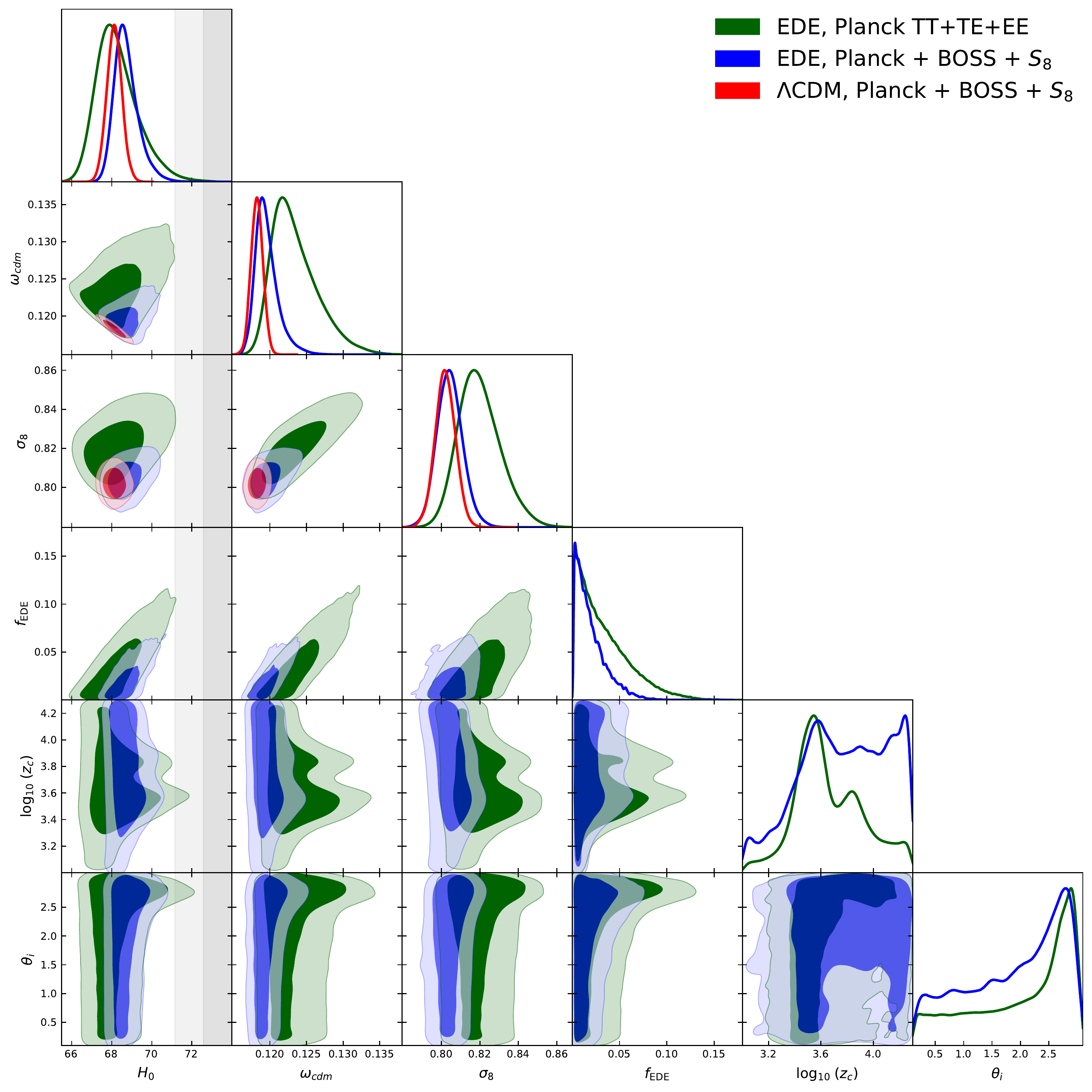}
    \caption{Cosmological parameter constraints from the joint {\it Planck} 2018 TT+TE+EE+low $\ell$+lensing + BOSS DR 12 (FS+BAO) + $S_8$ (DES+KV-450+HSC) likelihood.
    We also display the 
 constraints from the Planck 2018 primary CMB data alone (TT+TE+EE), obtained in \cite{Hill:2020osr}. 
    The SH0ES $H_0$
    measurement is shown in gray;
    the dark-shaded 
and light-shaded 
contours mark 
$68\%$ and $95\%$ confidence regions, respectively.
    }
    \label{fig:BOSS-S8}
\end{figure*}


\subsection{Full combination of CMB and LSS data}
\label{sec:S8}

 We now supplement our analysis with additional LSS data from DES-Y1, KV-450, and HSC via $S_8$. We note that these data sets are in mild ($\approx2.5\sigma$) tension with {\it Planck}+BOSS within the EDE model (see Table~\ref{table:params-P18+BOSS}).  However, we have found that each experiment (DES, KV-450, HSC) contributes roughly equally to the EDE constraints. Excluding any one of these experiments from our dataset makes the remaining combination consistent with {\it Planck}+BOSS and does not impact significantly our final conclusions. We use posterior sampling to estimate the result of a full combined analysis.

There are several important caveats in our weak lensing (WL) analysis, which should be borne in mind when interpreting our WL results.

First, we treat all LSS data sets as independent in our work.
The covariance between the different WL surveys that we use is small, as argued in Ref.~\cite{Hill:2020osr}.
This is due to a small sky overlap, different survey depths, and
different photo-$z$ calibration, 
which diminishes the impact of a possible common systematic.
As far as the covariance between BOSS and WL surveys are concerned, the sky overlap between BOSS
and the DES-Y1, KV450, and HSC footprints is only
$1\%$, $2\%$ and $1.5\%$
\cite{Lee:2019pfw,Troster:2019ean,Kondo:2019ind} of the total BOSS footprint, respectively, hence the cross-covariance is negligibly small as well.

The second caveat is regarding the  compression of the WL likelihoods; we approximate the WL surveys with a Gaussian prior on $S_8$. This procedure was validated in~\cite{Hill:2020osr} for DES both in the EDE and $\Lambda$CDM contexts. Since the likelihoods for other surveys are not available at the moment, we approximate the other WL likelihoods with the $S_8$ prior as well.  This practice should be accurate for these surveys as well; indeed, the $S_8$ estimates from HSC and KV-450 are more Gaussian than that of DES, for which the  Gaussian approximation to $S_8$ was shown to capture all relevant information in combination with Planck.  The $S_8$ compression, of course, misses some additional information contained in the WL surveys. However, since this information  was found to be negligible for DES Y1 (more precisely, to have minimal impact on the posterior distribution of EDE model parameters), we expect it should also be negligible for KV-450 and HSC, which have larger statistical errors and hence have less signal beyond $S_8$ than DES.

The third important caveat is that the KV-450 and HSC measurements are based on non-linear biasing models
similar to the that of the EFT.
This suggests that a 
more accurate analysis 
of the WL data should be carried out
in the EFT framework, with a common set of nuisance parameters for overlapping galaxy selections, as was done in Ref.~\cite{Troster:2019ean}. 
A proper joint analysis of the BOSS and WL data would break 
degeneracies
between the nuisance and cosmological parameters, which in turn
would strengthen the constraints compared to our analysis where 
the WL and BOSS galaxies 
are treated effectively as independent.
We adopt a similar conservative 
approach 
in our BOSS likelihood,
where we treat various data chunks 
independently on the basis that their selection
functions are slightly different,
although this difference 
is not statistically significant, see e.g.~\cite{DAmico:2019fhj} for the study of the NGC and SGC high-z samples. All in all, we leave the WL analysis within the EFT for future work. 
At this point, we present the results of our approximate treatment, which, nevertheless, is expected to be accurate enough for our purposes. 

The resulting parameter constraints from our analysis are given in Table~\ref{table:params-P18+BOSS-S8}, and the posterior distributions are shown in Fig.~\ref{fig:BOSS-S8} (the full triangle plot can be found in Appendix~\ref{app:full}). We find no evidence for the EDE component, but rather a $26\%$ improvement on the upper bound from the analysis with BOSS DR12 in the previous subsection.  We find $f_{\rm EDE} =0.019_{-0.019}^{+0.0040}$, consistent with null, with an upper bound $f_{\rm EDE}<0.053$ at $95\%$ CL.  Accordingly, there is very little constraining power on $\log_{10}(z_c)$ or $\theta_i$, as seen in the posteriors in Fig.~\ref{fig:BOSS-S8}.  We find the Hubble constant $H_0=68.73^{+0.42}_{-0.69}$ km/s/Mpc and $H_0=68.13 \pm 0.38$ km/s/Mpc in EDE and $\Lambda$CDM, respectively. The $\Lambda$CDM result is more than $4\sigma$ away from the SH0ES measurement (in units of the SH0ES standard deviation), while the EDE result is more than $3.5\sigma$ away from SH0ES,
indicating little prospect for resolving the Hubble tension within the EDE model.  From a Bayesian perspective, this tension implies that the SH0ES data set should not be combined with the others analyzed here, even in the broadened EDE parameter space. Overall, the constraints on the standard cosmological parameters are very similar in both the EDE and $\Lambda$CDM models, which suggests that current LSS data sets have almost saturated all possible channels to constrain the EDE model, i.e., almost all degeneracies between the EDE and standard cosmological parameters are broken.

Finally, we perform an additional test of our analyses, detailed in Appendix~\ref{app:prior}. It has been suggested \cite{Smith:2019ihp,Niedermann:2020dwg} that the null results of searches for EDE in combined data sets that do not include the SH0ES measurement are due to a failure of the Metropolis-Hastings algorithm to explore the EDE parameter space, stemming from parameter degeneracies that emerge in the $\Lambda$CDM limit, $f_{\rm EDE}\rightarrow 0$. This would suggest the existence of an as-yet hidden region of EDE parameter space that is able to accommodate both the local and cosmological data sets, despite being strongly excluded by the MCMC analyses heretofore performed. To address this, we perform an MCMC analysis with a lower bound $f_{\rm EDE}>0.04$, so as to force the sampler away from the $\Lambda$CDM limit of the EDE model, thus removing any parameter-space volume effects associated with degeneracies at $f_{\rm EDE} = 0$.  We use the same data set combination of $Planck$ 2018 + BOSS DR12 + $S_8$ as used in Fig.~\ref{fig:BOSS-S8}. The posterior distributions and parameter constraints are given in Fig.~\ref{fig:fEDE-prior} and Tab.~\ref{table:params-fEDE-prior}, respectively. Even in this artificial case, the posterior for $f_{\rm EDE}$ is centered at the lower bound of the prior. This strongly suggests that the combined cosmological dataset does not favor any non-vanishing amount of EDE, regardless of parameter-space volume effects. Further details of this analysis, as well as a frequentist $\chi^2$ comparison, are given in Appendix~\ref{app:prior}.

\section{Forecast EDE constraints with future LSS data}

The next generation of LSS experiments will dramatically increase the volume and precision of LSS data, and in particular, \textit{Euclid} \cite{Amendola:2016saw}, DESI \cite{Levi:2019ggs}, WFIRST/Roman \cite{2019arXiv190205569A}, and Vera Rubin Observatory \cite{Ivezic:2008fe} (formerly LSST), will provide an abundance of data across a range of redshifts. It is expected that these new data sets will significantly improve parameter inferences for the $\Lambda$CDM model and its extensions.

\begin{table*}[htb!]
 Constraints from \emph{Planck} 2018  + mock \textit{Euclid} 
and \emph{Planck} 2018  + BOSS DR12 data
 \vspace{2pt} \\
  \centering
  \begin{tabular}{|l|c|c|}
    \hline\hline Parameter & EDE ($n=3$), BOSS &  EDE ($n=3$), {\it Euclid}\\ \hline \hline

    {\boldmath$\ln(10^{10} A_\mathrm{s})$}  & $3.047 \, (3.049) \, \pm 0.014$& $3.043\, (3.048)_{-0.0057}^{+0.0070}$ \\

    {\boldmath$n_\mathrm{s}$}  & $0.9696 \, (0.9717)^{+0.0046}_{-0.0068}$ & $0.9641 (0.9618)_{-0.003}^{+0.0031}$\\

    {\boldmath$100\theta_\mathrm{s}$} & $1.04172 \, (1.04126) \, \pm 0.00032$ & $1.042 (1.042)_{-0.00029}^{+0.00027} $ \\

    {\boldmath$\Omega_\mathrm{b} h^2$} &  $0.02255 \, (0.02245) \, \pm 0.00018$ & $0.02237 (0.02232)_{-0.00011}^{+0.00011} $  \\

    {\boldmath$\Omega_\mathrm{cdm} h^2$}  & $0.1215 \, (0.1243)^{+0.0013}_{-0.0029}$ & $ 0.1204 (0.1204)_{-0.00055}^{+0.00042}$ \\

    {\boldmath$\tau_\mathrm{reio}$} & $0.05533 \, (0.05433)_{-0.0075}^{+0.0069}$ & $0.05554 (0.5335)_{-0.004}^{+0.0041}$ \\

    {\boldmath$\mathrm{log}_{10}(z_c)$} & $3.71 \, (3.52)^{+0.26}_{-0.33}$& $3.46 \, (3.47)^{+0.17}_{-0.15}$  \\

    {\boldmath$f_\mathrm{EDE} $}& $< 0.072 \, (0.047)$  & $<0.012\,(0.0023)$ \\

    {\boldmath$\theta_i$}  & $2.023(2.734)_{-0.34}^{+1.1} $& $2.634(2.73)_{-0.069}^{+0.47}$\\

    \hline

    $H_0 \, [\mathrm{km/s/Mpc}]$ & $68.54 \, (68.83)^{+0.52}_{-0.95}$ & $67.5\, (67.26)_{-0.22}^{+0.19}$ \\

    $\Omega_\mathrm{m}$ & $0.3082 \,(0.312)_{-0.0057}^{+0.0056}$& $0.3149\,(0.317)_{-0.0023}^{+0.0022}$  \\

    $\sigma_8$& $0.8127 \, (0.8195)_{-0.0091}^{+0.0072}$ & $0.8104\, (0.8115)_{-0.0021}^{+0.0022}$  \\
    
    $S_8$  & $0.8237 \, (0.8275)\, \pm 0.011$& $0.83038 \, (0.83209) \, \pm 0.0032$ \\
    \hline
  \end{tabular} 
  \caption{The mean (best-fit) $\pm1\sigma$ constraints on the cosmological parameters in the EDE scenario with $n=3$, as inferred from the combination of BOSS FS+BAO and \emph{Planck} 2018  TT+TE+EE + lensing data 
  and from the combination 
  of the same CMB data with the mock \textit{Euclid} likelihood.
  The upper limit on $f_{\rm EDE}$ is quoted at 95\% CL.}
  \label{table:params-P18+euclid}
\end{table*}

\begin{figure*}[ht!]
\begin{center}
\includegraphics[width=0.8\textwidth]{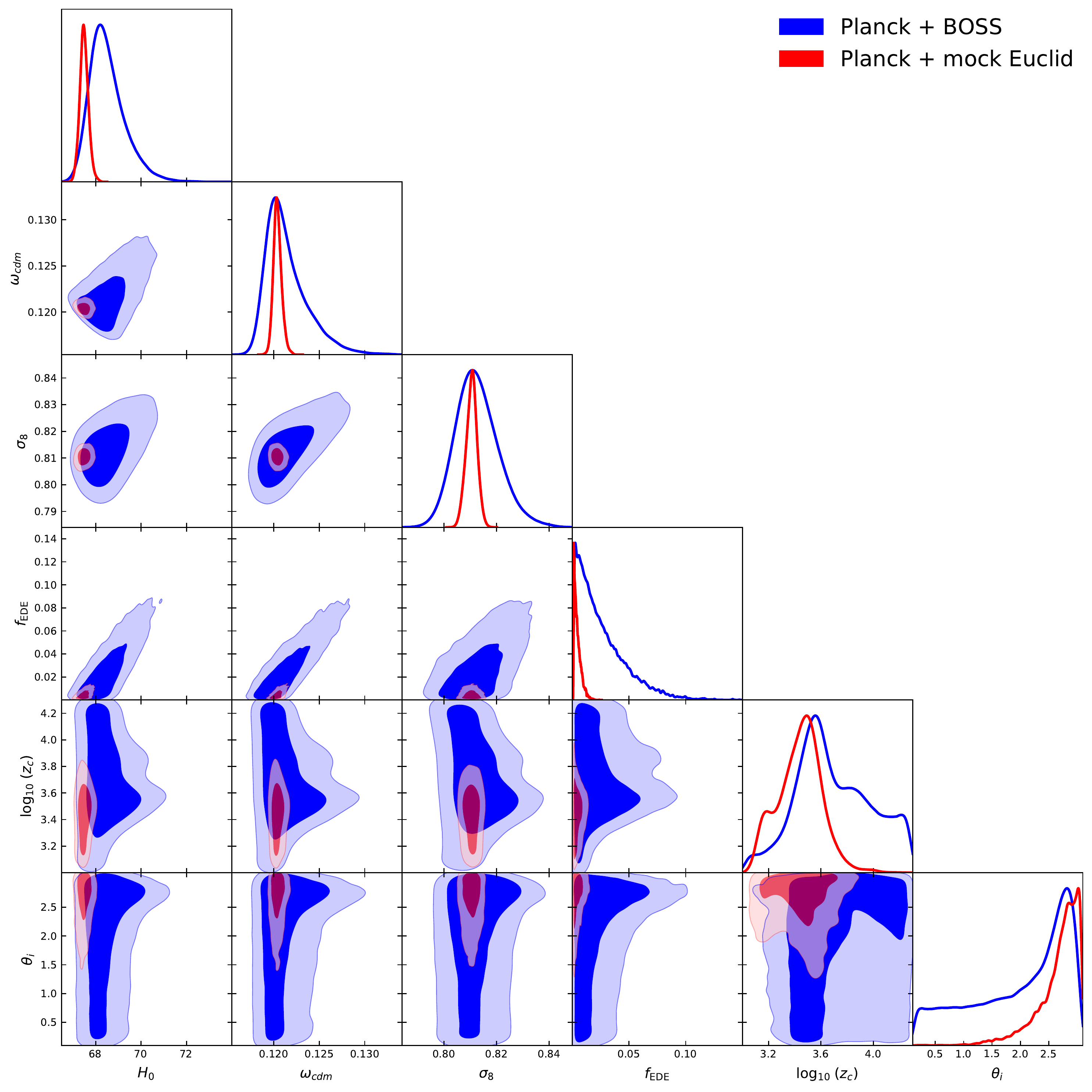}
\end{center}
\caption{Posterior distributions for the
parameters extracted from the joint {\it Planck} 2018 TT+TE+EE+low $\ell$+lensing + mock \textit{Euclid}/DESI likelihood, compared to those from {\it Planck} + BOSS data.
\label{fig:planck-euclid} } 
\end{figure*}

 We have thus far demonstrated that current data sets severely constrain the EDE extension of $\Lambda$CDM, imposing a 95\% CL upper bound $f_{\rm EDE}<0.053$ (see Table \ref{table:params-P18+BOSS-S8}). It is reasonable to expect that the next generation of experiments will further improve on this bound, or aid in detecting a small EDE component if it is indeed present in the universe. To quantify this, we now perform an EDE sensitivity forecast for next-generation LSS experiments. For concreteness, we focus on \textit{Euclid} \cite{Amendola:2016saw}, which is expected to measure the redshifts of $5 \cdot 10^7$ galaxies in the interval $0.5< z <2.1$. We expect the results for DESI would be very similar because the two surveys have comparable characteristics (note that \textit{Euclid} will also perform photometric imaging, however).

We perform an MCMC analysis of the EDE model with a combined data set comprised of a forecast mock likelihood for \textit{Euclid} and the final {\it Planck} 2018 CMB data. We construct the forecast \textit{Euclid} likelihood assuming the best-fit base $\Lambda$CDM model found by {\it Planck} 2018. This methodology is very close to a Fisher forecast, but it is free from the assumptions of the Fisher approximation. In particular, it is accurate in the case of a non-Gaussian posterior distribution, which is especially important for the EDE model.

Our goal is to pinpoint the additional constraining power due to \textit{Euclid} in the EDE analysis with combined CMB+LSS data. To this end, we construct a forecast \textit{Euclid} likelihood assuming the null hypothesis ($f_{\rm EDE}=0$) and perform a sensitivity forecast for $f_{\rm EDE}$. This approach is consistent with the fact that the current {\it Planck}+LSS data show no evidence for the EDE scenario. We additionally note that the constraining power of the real {\it Planck} 2018 TT+TE+EE+low $\ell$ + lensing likelihood is significantly stronger than publicly-available mock CMB likelihoods, such as the \texttt{fake\_planck\_realistic} likelihood included in \texttt{Monte-Python v-3.1} \cite{Brinckmann:2018cvx}, which does not include the 217 GHz channel present in the real {\it Planck} data. We further note that present (real) TT+TE+EE {\it Planck} data alone already excludes most of the region of EDE parameter space relevant to the Hubble tension (see Table I of \cite{Hill:2020osr}). For these reasons, as well as to better isolate the impact of \textit{Euclid}, we do not utilize a mock CMB likelihood for this analysis, and instead rely on real CMB data.

We use the mock \textit{Euclid} redshift-space power spectrum likelihood introduced in \cite{Chudaykin:2019ock}, with the survey specification of Ref.~\cite{Pozzetti:2016cch}. The mock power spectra are generated using the forecast for the density distribution of emission-line galaxies (ELGs) across 8 non-overlapping redshift bins of width $\Delta z = 0.2$ spanning the range $0.5<z<2$. Assuming the sky fraction of the survey $f_{\rm sky}=0.3636$, this yields a total volume of $\sim 70~$(Gpc/$h)^3$, which is roughly 10 times bigger than the volume of the BOSS survey.

Our mock likelihood is based on the same  non-linear model for matter clustering, bias, redshift-space distortions and baryonic effects as the one used for the BOSS data analysis. However, there are two minor differences 
that should be mentioned. First, we do not scan over the next-to-leading order fingers-of-God counterterm $\tilde{c}$. This term was omitted because the ELGs are expected to be less affected by fingers-of-God~\cite{Chudaykin:2019ock} than the Luminous Red Galaxies observed by BOSS. Second, we include the hexadecapole moment in our forecast Euclid likelihood. This moment is accompanied by 
an additional $k^2-$ counterterm. All in all, our theoretical model captures various non-linear effects by 7 nuisance parameters in every redshift bin, which totals to 56 LSS-related  parameters to be marginalized over. We do not assume any priors on the nuisance parameters in order to keep the analysis more conservative. Our mock likelihood also includes the two-loop theoretical error covariance \cite{Baldauf:2016sjb}, which automatically implements realistic data cuts. Note that unlike the baseline analysis of Ref.~\cite{Chudaykin:2019ock}, we do not include the bispectrum data, which could improve the constraints on the EDE scenario even further. 

The parameter constraints from the fit to the combined real {\it Planck} likelihood and forecast \textit{Euclid} likelihood are given in Table~\ref{table:params-P18+euclid}, and the posterior distributions are shown in Fig.~\ref{fig:planck-euclid}. We find that the EDE component is constrained to peak at less than $1.2\%$ of the energy density of the universe; we find $f_{\rm EDE}<0.012$ at 95\% CL. This is a factor of 6 improvement over the constraints from {\it Planck} 2018 with BOSS DR12 (Table \ref{table:params-P18+BOSS}). We find a more modest improvement on the Hubble constant, with an error that is decreased from $\sigma(H_0)\simeq 0.70$ km/s/Mpc to $\sigma(H_0)\simeq 0.20$ km/s/Mpc.

As in the $\Lambda$CDM case discussed in Refs.~\cite{Chudaykin:2019ock,Ivanov:2019hqk}, the primary gain comes from better measurements of $\omega_{cdm}$ from the power spectrum shape, $H_0$ from the BAO, and $f\sigma_8$, probed through RSD. These improvements are comparable to the ones obtained for $\Lambda$CDM in Ref.~\cite{Chudaykin:2019ock} and roughly correspond to the increase in the number of measured modes, $\sqrt{V_{\rm Euclid}/V_{\rm BOSS}}\sim 3$. Our results follow the trend seen in the analysis of the \textit{Planck}+BOSS+$S_8$ data: the constraints on the parameters of the EDE scenario are very similar to the $\Lambda$CDM ones, which indicates that all detection channels for the EDE are nearly exhausted after the addition of the LSS data.

\section{Discussion and Conclusions}
\label{sec:Discussion}

The EDE scenario is a potentially compelling early-universe resolution to the persistent and increasingly significant disagreement in late vs.~early universe inferences of the Hubble constant. The EDE model successfully decreases the comoving sound horizon, allowing for a fixed angular size of the sound horizon even for an increased $H_0$. Combined with accompanying shifts in the other standard $\Lambda$CDM parameters, e.g., the physical dark matter density, this can provide a good fit to the {\it Planck} 2018 CMB temperature and polarization data for $H_0$ values in near-agreement with SH0ES. However, as emphasized in \cite{Hill:2020osr}, these parameter shifts are in tension with other cosmological data sets, and in particular, LSS data. This comes at a time when LSS data, in combination with a BBN prior on the baryon density, provides a CMB-independent early universe measurement of $H_0$ \cite{Abbott:2017smn,Ivanov:2019hqk,DAmico:2019fhj,Philcox:2020vvt}, that is consistent with the value inferred from {\it Planck} 2018 CMB data.

Past claims of evidence for the EDE scenario (e.g.,~\cite{Poulin:2018cxd,Smith:2019ihp}) were based on the \textit{Planck} CMB data combined with several external datasets, such as SH0ES, BAO, supernovae, and $f\sigma_8$ measurements from RSD.  Crucially, SH0ES was used in the joint analysis without first checking whether the $H_0$ posterior from the non-SH0ES data sets was statistically consistent with the SH0ES measurement.  As shown in~\cite{Hill:2020osr} and in this paper, this is not the case.  Thus, from a Bayesian perspective, one should not analyze SH0ES in tandem with the other cosmological data sets.  Moreover, the datasets considered in previous claimed EDE detections are not complete.  First, they exclude photometric galaxy clustering and weak lensing data. Second, they rely on a simplified ``compressed'' redshift-space galaxy power spectrum likelihood that ignores the matter power spectrum shape information and implicitly
assumes standard early-universe physics.

The impact of the galaxy clustering and 
weak lensing data
on the EDE constraints was recently studied in 
Refs.~\cite{Hill:2020osr}
and \cite{Chudaykin:2020acu}.
Hill et al.~(2020)~\cite{Hill:2020osr} first
showed that the primary CMB data alone does not reveal significant evidence 
for the EDE model.
Moreover, the constraints on the EDE model strengthen after taking into account the data from photometric surveys.
The ``walking barefoot'' analysis 
of Ref.~\cite{Hill:2020osr},
based on all available cosmological datasets  \textit{without} SH0ES, yielded an upper limit $f_{\rm EDE}<0.060$ ($95\%$CL), significantly lower than the value $f_{\rm EDE}\approx 0.1$ needed to resolve the Hubble tension. Thus, the addition of the LSS data rules out the EDE model as a resolution to the Hubble tension.

Chudaykin et al.~(2020)~\cite{Chudaykin:2020acu} claimed that 
the photometric LSS data 
does not rule out the EDE model 
if the $\ell>1000$ region of the {\it Planck} power spectra are discarded
and replaced with the SPTPol measurements \cite{Henning:2017nuy}.
This was motivated by the presence of the so-called ``lensing anomaly''
in the {\it Planck} high-$\ell$ data.
The significance of this anomaly is $2.8\sigma$ \cite{Aghanim:2018eyx}, which still makes it compatible with a statistical fluctuation, and no systematic has been identified as a culprit despite significant dedicated analysis~\cite{Addison:2015wyg,Aghanim:2016sns}. Thus, we believe that the presence of this mild tension does not give a sufficiently strong reason to discard the {\it Planck} high-$\ell$ data, which has more 
statistical power than the SPTPol measurement.  It is also worth noting that $\Lambda$CDM does not provide a very good fit to the SPTPol power spectra (PTE = 0.017), and there are mild internal parameter tensions within the SPTPol data set (see Sec.~8 of~\cite{Henning:2017nuy}).

In this paper, we  have investigated if the addition of the full BOSS galaxy power spectrum likelihood, with all cosmological parameters varied, can change the conclusions of previous analyses based on a compressed ``official'' version of this likelihood. In particular, while past analyses of EDE have relied on a direct application of the official RSD BOSS likelihood, which can be seen as constructed from the full BOSS likelihood by assuming standard early-universe physics as implemented through priors on the shape of the power spectrum, in this work we have instead used the full BOSS likelihood, with the power spectrum computed in a self-consistent manner as described in Sec.~\ref{sec:EDExLSS}. This removes any uncertainty as to the validity of LSS constraints on the EDE model.

The full constraining power of redshift-space galaxy clustering data can be accessed through the application of the EFT of LSS~\cite{Baumann:2010tm,Carrasco:2012cv}. This is evidenced through \cite{Ivanov:2019hqk,DAmico:2019fhj,2020A&A...633L..10T,Philcox:2020vvt}, which find precise constraints on cosmological parameters while relying only on the final data release of BOSS. Motivated by this, and the results of \cite{Hill:2020osr}, here we have applied the EFT to the EDE model, and computed parameter constraints from LSS data. The EDE model can be succinctly parametrized by three parameters: $f_{\rm EDE}$, $z_c$, and $\theta_i$, which correspond to the peak energy density as a fraction of the universe, the timing of this peak, and the initial condition for the EDE field. See Fig.~\ref{fig:EDEn3} for an illustrative example. Previous analyses have shown $f_{\rm EDE} \approx 10\%$ can resolve the Hubble tension \cite{Poulin:2018cxd,Smith:2019ihp}. Our analyses using the EFT lead to stringent upper bounds on $f_{\rm EDE}$, which are incompatible with EDE as a resolution to the Hubble tension. 

We find that BOSS data, including the complete full-shape and BAO likelihoods, in combination with {\it Planck} 2018 data, lead to a upper bound $f_{\rm EDE }< 0.072$ at 95\% confidence. A $68\%$ confidence limit yields $f_\mathrm{EDE} = 0.025_{-0.025}^{+0.0061}$, consistent with zero.  We find a value for the Hubble constant $H_0=68.54^{+0.52}_{-0.95}$ km/s/Mpc, in $3.6\sigma$ tension with the SH0ES measurement  ($H_0=74.03\pm1.42$ km/s/Mpc). Supplemented with additional LSS data in the form of an $S_8$ prior corresponding to the measurements of DES-Y1, KV-450, and HSC (a procedure that was validated for the EDE model in \cite{Hill:2020osr}), we find an upper bound $f_{\rm EDE} < 0.053$, with a $68\%$ confidence limit $f_\mathrm{EDE} = 0.019_{-0.019}^{+0.0040}$. We find the Hubble constant $H_0=68.73^{+0.42}_{-0.69}$ km/s/Mpc, again discrepant with the SH0ES measurement at $3.6\sigma$ significance.

Overall, the constraints obtained in this work are similar to those of the ``walking barefoot'' (no-SH0ES) analysis from Hill et. al. (2020) \cite{Hill:2020osr}. Both analyses rule out the EDE model as a plausible resolution of the Hubble tension. We note that our conclusions are different from the recent results of~\cite{Niedermann:2020dwg}. However, the analysis in~\cite{Niedermann:2020dwg} does not use the full-shape BOSS likelihood and implements a non-standard method in the $f_{\rm EDE}\to 0$ limit, in which some EDE parameters are held fixed rather than varied in the MCMC. It would be interesting to see if the discrepancy in results prevails once the same pipeline is used.

Our results indicate that addition of the full shape information from the BOSS galaxy power spectrum significantly improves the constraints on the EDE scenario compared to both the primary \textit{Planck}-only and \textit{Planck} + standard FS BOSS results. In particular, it allows us to rule out (at $95\%$ CL) the region of the EDE parameter space that addresses the Hubble tension in a combined analysis with the \textit{Planck} data. This can be contrasted with the standard BAO/RSD BOSS likelihood based on the approximate BAO+$f\sigma_8$ parametrization, which (a) does not noticeably improve the \textit{Planck}-only constraints and (b) is moderately compatible with a significant amount of EDE. This comparison is detailed in Appendix~\ref{app:fs8}.

The importance of a consistent analysis of the galaxy clustering data will increase even further in the era of future surveys. Indeed, the coming decade will see the deployment of a new generation of LSS experiments, e.g.,~\textit{Euclid}~\cite{Amendola:2016saw}, DESI~\cite{Levi:2019ggs}, WFIRST/Roman~\cite{2019arXiv190205569A}, and Vera Rubin Observatory~\cite{Ivezic:2008fe}, each with the explicit aim of doing precision cosmology with LSS. In light of this, we have performed a sensitivity forecast of next-generation LSS experiments to EDE. We have constructed a mock likelihood for {\it Euclid} assuming a non-observation of EDE, with a fiducial cosmology given by $\Lambda$CDM with the best-fit parameters of {\it Planck} 2018. Taken in conjunction with current {\it Planck} data, we obtain an upper bound $f_{\rm EDE}<0.012$ at 95\% CL, which would constrain the fraction of the universe in the EDE component, at the peak of its evolution, to be less than $1.2\%$.

These results indicate a bleak outlook for EDE as a resolution to the Hubble tension. On the other hand, the results are highly encouraging for the use of the EFT of LSS as a probe of physics beyond the standard model, which will be crucial for next-generation LSS experiments. Experiments such as {\it Euclid}, DESI, WFIRST, and VRO can be  expected to tightly constrain not only the standard $\Lambda$CDM parameters, but also significant model extensions, such as EDE. While the prospect of EDE as a Hubble tension resolution looks increasingly unlikely, this result serves as motivation for further theoretical exploration to find a model that successfully yields a high $H_0$ value when fit to the wealth of precision cosmological data available today.\\

\acknowledgments
The authors thank
Anton Chudaykin, Jo Dunkley, Shaun Hotchkiss, 
Oliver Philcox,
Marcel Schmittfull, and Bill Wright for helpful discussions, as well as the Scientific Computing Core staff at the Flatiron Institute for computational support.  JCH thanks the Simons Foundation for support.
The work of MI on parameter constraints for the EDE model is supported by the Russian Science Foundation (grant 18-12-00258).
EM is supported in part by a Banting Fellowship from the government of Canada.
MZ is supported by NSF grants AST1409709, PHY-1521097 and PHY-1820775 the Canadian
Institute for Advanced Research (CIFAR) program on
Gravity and the Extreme Universe and the Simons Foundation Modern Inflationary Cosmology initiative.

\bibliographystyle{JHEP}
\bibliography{EDE-LSS-refs}

\clearpage

\onecolumngrid

\pagebreak
\appendix

\section{Full triangle plots}
\label{app:full}

\begin{figure*}[ht!]
\begin{center}
\includegraphics[width=1.0\textwidth]{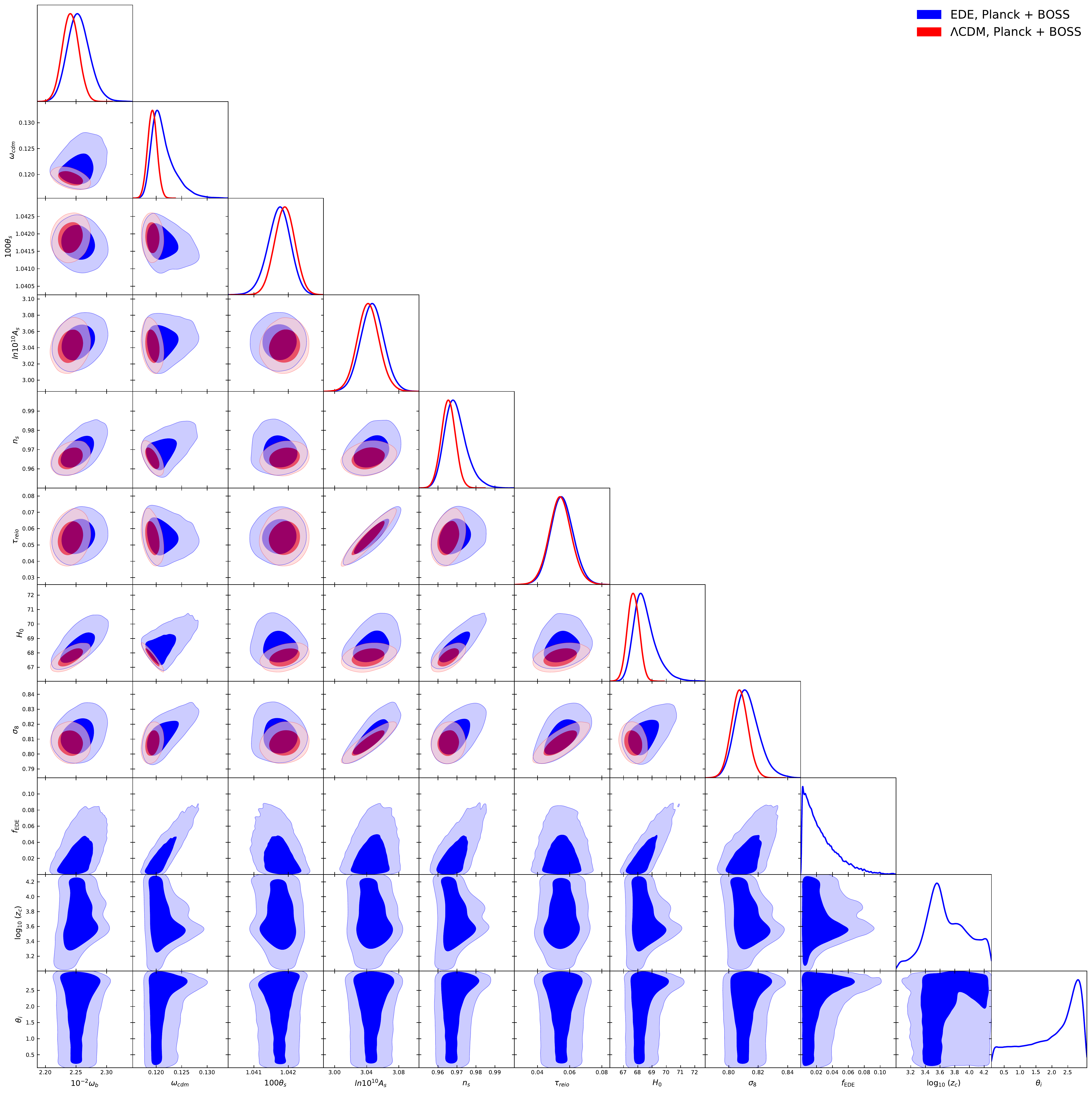}
\end{center}
\caption{Posterior distributions for the
parameters extracted from the joint \emph{Planck} 2018 TT+TE+EE + low $\ell$ + lensing 
+ BOSS DR12 (FS+BAO) likelihood.
\label{fig:planck+BOSS-big} } 
\end{figure*}

\begin{figure*}[h]
\centering
 \includegraphics[width=\textwidth]{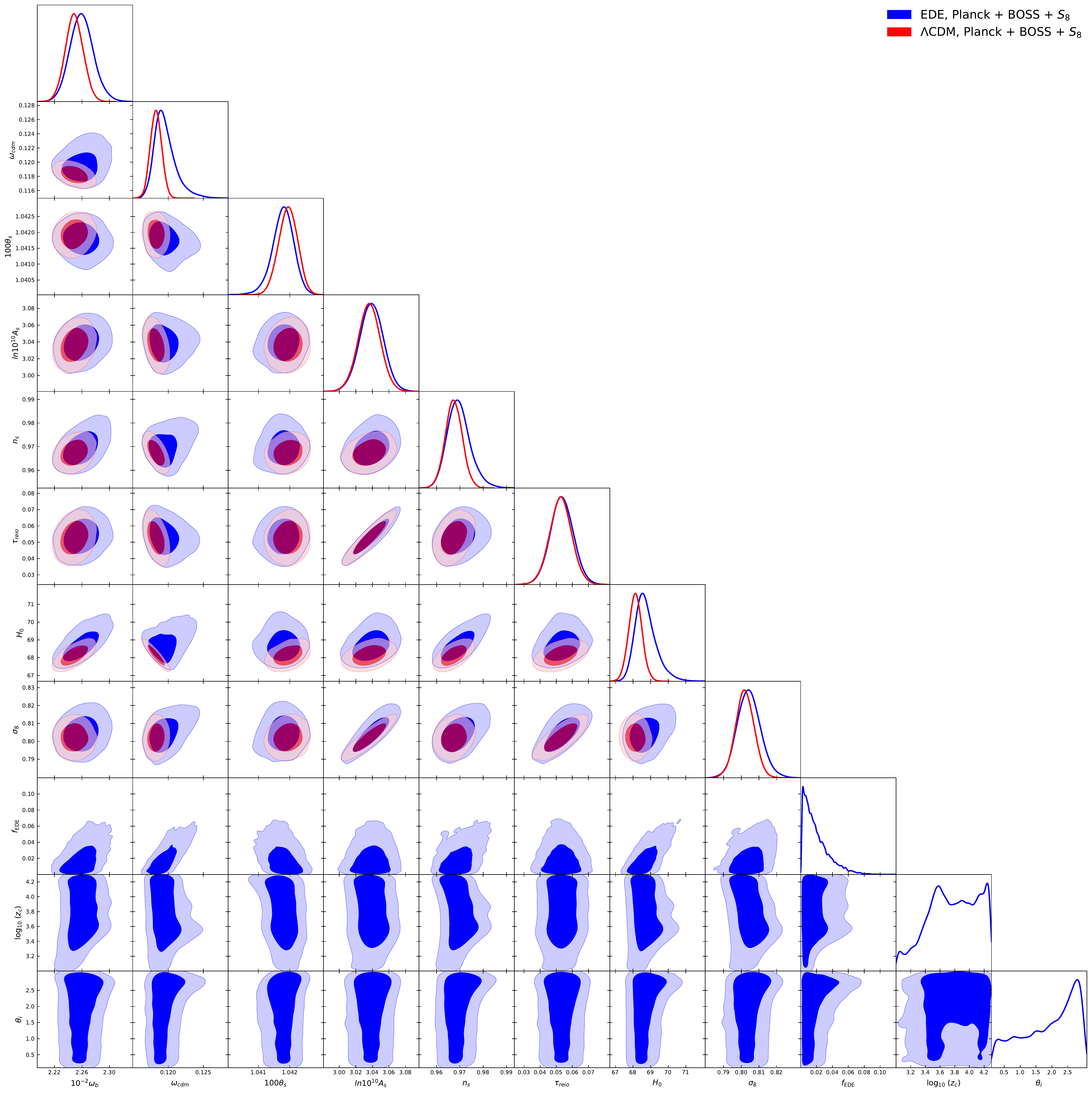}
    \caption{Cosmological parameter constraints from the joint \emph{Planck} 2018 TT+TE+EE + low $\ell$ + lensing + BOSS DR12 + $S_8$ likelihood.  
    }
    \label{fig:BOSS-S8-big}
\end{figure*}

\begin{figure*}[h!]
\begin{center}
\includegraphics[width=1.0\textwidth]{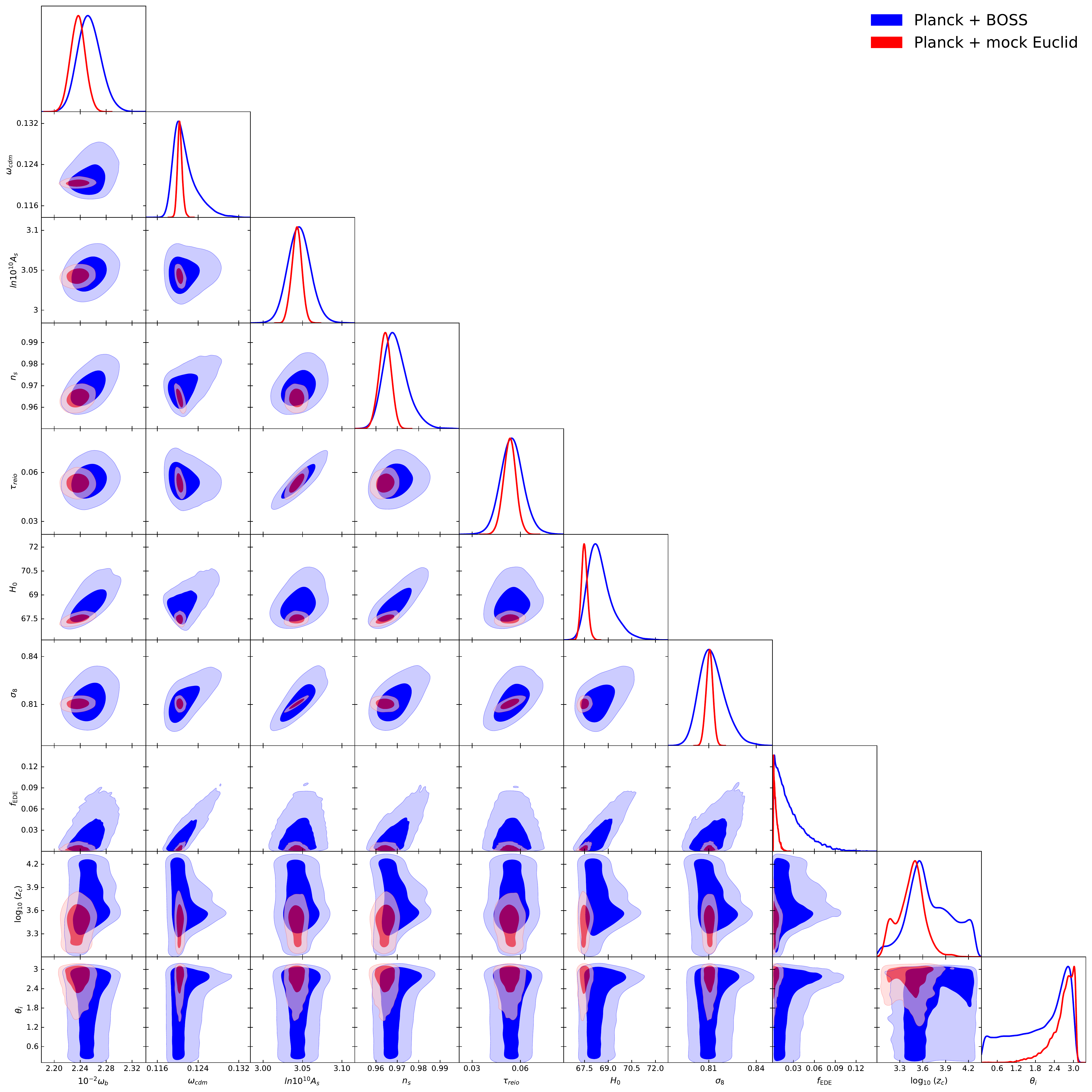}
\end{center}
\caption{ Posterior distributions for the
EDE model
parameters extracted from the joint \emph{Planck} 2018 TT+TE+EE + low $\ell$ + lensing + mock \emph{Euclid}/DESI likelihood.
\label{fig:planck-euclid-big} } 
\end{figure*}

In this appendix we present full
triangle plots for the cosmological parameters of the EDE model, considering the following combinations of data: \textit{Planck} + BOSS in Fig.~\ref{fig:planck+BOSS-big},
\textit{Planck} + BOSS + $S_8$ in 
Fig.~\ref{fig:BOSS-S8-big}, and
\textit{Planck} + mock {\it Euclid}
in Fig.~\ref{fig:planck-euclid-big}.
In the first two cases of real data,
we also show the $\Lambda$CDM constraints obtained from the same 
datasets.
Additionally, in Tables~\ref{table:params-P18+BOSS-2tailed}
and~\ref{table:params-P18+BOSS-S8-2tailed}
we present alternative estimates of the confidence limits
for the two analyses of the real data
using the equal-tail method (see discussion near the beginning of Sec.~\ref{sec:constraints}).

\begin{table*}[htb!]
Constraints from \emph{Planck} 2018 data  + BOSS DR12 \vspace{2pt} \\
  \centering
  \begin{tabular}{|l|c|c|}
    \hline\hline Parameter &$\Lambda$CDM~~&~~~EDE ($n=3$) ~~~\\ \hline \hline

    {\boldmath$\ln(10^{10} A_\mathrm{s})$} & $3.042 \, (3.034) \, \pm 0.014$ & $3.047 \, (3.049) \, \pm 0.014$ \\

    {\boldmath$n_\mathrm{s}$} & $0.9655 \, (0.9655) \, \pm 0.0037 $ & $0.9694 \, (0.9717)\pm 0.0055$ \\

    {\boldmath$100\theta_\mathrm{s}$} & $1.04185 \, (1.04200) \, \pm 0.00029 $ & $1.04173 \, (1.04126) \, \pm 0.00032$\\

    {\boldmath$\Omega_\mathrm{b} h^2$} & $0.02241 \, (0.02233) \, \pm 0.00013 $ &  $0.02254 \, (0.02245) \, \pm 0.00018$ \\

    {\boldmath$\Omega_\mathrm{cdm} h^2$} & $ 0.1192\, (0.1191)\pm 0.00091$ & $0.1214 \, (0.1243) \pm 0.0033$ \\

    {\boldmath$\tau_\mathrm{reio}$} & $0.05428 \, (0.0503)\pm 0.0068 $ & $0.05520 \, (0.05433)\pm 0.0070$\\

    {\boldmath$\mathrm{log}_{10}(z_c)$} & $-$ & $3.70 \, (3.52)^{+0.34}_{-0.27}$ \\

    {\boldmath$f_\mathrm{EDE} $} & $-$ & $< 0.072 \, (0.047)$\\

    {\boldmath$\theta_i$} & $-$ & $1.995(2.734)_{-0.27}^{+0.34} $\\

    \hline

    $H_0 \, [\mathrm{km/s/Mpc}]$ & $67.68 \, (67.56) \, \pm 0.42$ & $68.51 \, (68.83)^{+0.75}_{-0.73}$ \\

    $\Omega_\mathrm{m}$ & $0.3107 \, (0.3112)\pm 0.0055$ & $0.3082 \,(0.3120)\pm 0.0057$ \\

    $\sigma_8$ & $0.8074 \, (0.8039) \, \pm 0.0056$ & $0.8126 \, (0.8195)_{-0.0079}^{+0.0080}$ \\
    $S_8$ &      $0.822 \, (0.819) \, \pm 0.010$ 
    & 
       $0.824 \, (0.827)\, \pm 0.011$ 
    \\
    \hline
  \end{tabular} 
  \caption{The mean (best-fit) $\pm1\sigma$ constraints on the cosmological parameters in $\Lambda$CDM and in the EDE scenario with $n=3$, as inferred from the combination of BOSS FS+BAO and \emph{Planck} 2018 TT+TE+EE + low $\ell$ + lensing data. The upper limit on $f_{\rm EDE}$ is quoted at 95\% CL. These results use the equal-tail method discussed in Sec.~\ref{sec:constraints}.}
  \label{table:params-P18+BOSS-2tailed}
\end{table*}

\begin{table*}[htb!]
Constraints from \emph{Planck} 2018 data  + BOSS DR12  + $S_8$ from DES+KV-450+HSC\vspace{2pt} \\
  \centering
  \begin{tabular}{|l|c|c|}
    \hline\hline Parameter &$\Lambda$CDM~~&~~~EDE ($n=3$) ~~~\\ \hline \hline

    {\boldmath$\ln(10^{10} A_\mathrm{s})$} & $3.036 \, (3.039) \, \pm 0.014$ & $3.038 \, (3.034) \, \pm 0.014$ \\

    {\boldmath$n_\mathrm{s}$} & $0.9674 \, (0.9727) \, \pm 0.0037 $ & $0.9696 \, (0.9621)^{+0.0045}_{-0.0047}$ \\

    {\boldmath$100\theta_\mathrm{s}$} & $1.041945 \, (1.041966) \, \pm 0.00030 $ & $1.04178 \, (1.04176) \, \pm 0.00035$\\

    {\boldmath$\Omega_\mathrm{b} h^2$} & $0.02249 \, (0.02273) \, \pm 0.00013 $ &  $0.02259 \, (0.022433)\pm 0.00017$ \\

    {\boldmath$\Omega_\mathrm{cdm} h^2$} & $ 0.1182\, (0.1157)\pm 0.00081$ & $0.1196 \, (0.1195)^{+0.0016}_{-0.0015}$ \\

    {\boldmath$\tau_\mathrm{reio}$} & $0.052726 \, (0.05911)\pm 0.0067 $ & $0.05349 \, (0.05211)_{-0.0068}^{+0.0069}$\\

    {\boldmath$\mathrm{log}_{10}(z_c)$} & $-$ & $3.77 \, (4.24)^{+0.39}_{-0.35}$ \\

    {\boldmath$f_\mathrm{EDE} $} & $-$ & $< 0.0526 \, (0.0115)$\\

    {\boldmath$\theta_i$} & $-$ & $1.91(1.55)_{-1.1}^{+0.9} $\\

    \hline

    $H_0 \, [\mathrm{km/s/Mpc}]$ & $68.13 \, (69.28) \, \pm 0.38$ & $68.73 \, (67.92)^{+0.55}_{-0.56}$ \\

    $\Omega_\mathrm{m}$ & $0.3046 \, (0.2859)\pm 0.0049$ & $0.3024 \,(0.3091)\pm 0.0050$ \\

    $\sigma_8$ & 
    $0.8020 \, (0.7947) \, \pm 0.0053$ & $0.8044 \, (0.8023)\pm 0.0062$ \\
    
    $S_8$ & 
    $0.8082 \, (0.7810) \, \pm 0.0086$ 
    & 
    $0.8075 \, (0.8143)\, \pm 0.0092$ 
    \\
    \hline
  \end{tabular} 
  \caption{
  The mean (best-fit) $\pm1\sigma$ constraints on the cosmological parameters in $\Lambda$CDM and in the EDE scenario with $n=3$, as inferred from the combination of BOSS FS+BAO, \emph{Planck} 2018 TT+TE+EE + low $\ell$ + lensing, and DES+KV-450+HSC data.  The upper limit on $f_{\rm EDE}$ is quoted at 95\% CL. These results use the equal-tail method discussed in Sec.~\ref{sec:constraints}.}
  \label{table:params-P18+BOSS-S8-2tailed}
\end{table*}

\clearpage
\pagebreak 
\section{Likelihood profile for $f_{\rm EDE}$}
\label{app:like}

\begin{figure*}[h!]
\begin{center}
\includegraphics[width=0.49\textwidth]{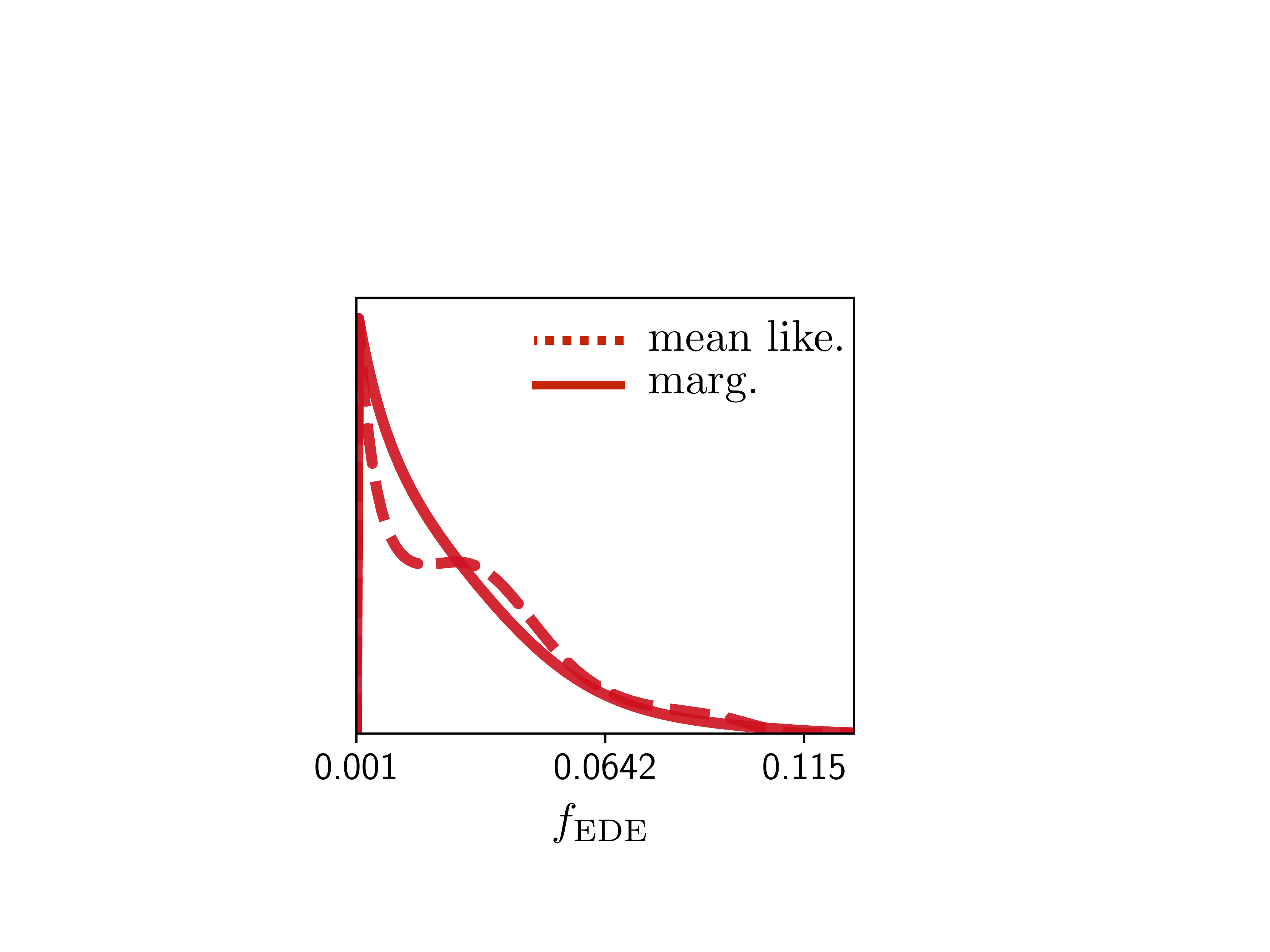}
\end{center}
\caption{ 
The marginalized 
1d posterior distribution (solid)
and the mean likelihood profile 
(dashed)
for $f_{\rm EDE}$, extracted from the \textit{Planck}+BOSS chains.
The normalization is such that the maximum of the distribution is equal to $1$.
\label{fig:like} } 
\end{figure*}

In Fig.~\ref{fig:like} we present 
the average 
likelihood curve extracted from our MCMC samples of the \textit{Planck}+BOSS analysis
for different 
values of $f_{\rm EDE}$. 
This is a standard output
of the \texttt{Monte Python} code, obtained 
as follows. 
First, we compute a grid of $f_{\rm EDE}$ in the range [0.001,0.12] with 20 equally-spaced steps. 
Then, for every $f_{\rm EDE}$-bin,
we find all MCMC 
steps with $f_{\rm EDE}$ values within the bin
and takes an average of the likelihood over all these steps.
Thus, we take into account all parameters that were varied in the MCMC chains. 

The difference of our procedure from the  best-fit test is that we show the likelihood profile constructed by taking the means of the $\chi^2$-samples
from our chains that have a fixed value of $f_{\rm EDE}$, and not the actual best-fits. In other words, we present a smoothed
version of the best-fit $\chi^2$, and it has several advantages over the best-fits. First, the smoothing down-weights the parameters that have to be fine-tuned to produce a good fit. This is especially important in the context of the EDE, which has 3 extra parameters compared to the base $\Lambda$CDM. Second, it is much more computationally cheap to extract the mean likelihood profile from the chains 
than to compute best-fits. 
Indeed, the calculation of the actual best-fit $\chi^2$ is difficult because of the presence of a large amount of nuisance 
parameters, whose distribution is often flat. 
Using the average likelihood profiles to quantify the goodness of fit and the prior-volume effects is the standard practice 
in the CMB data analyses~\cite{LewisBridle2002,Audren:2012wb,Planck2015likelihood}, and we believe that it is suitable to our purposes as well.

It is evident
that this profile is a monotonically 
decreasing function 
of $f_{\rm EDE}$, which 
implies that the EDE model 
does not improve the fit 
over $\Lambda$CDM
on average.
We find that the typical difference in the 
effective 
$\chi^2$-statistics between 
$\Lambda$CDM ($f_{\rm EDE}\to 0$)
and $f_{\rm EDE}=0.1$
is $\Delta \chi^2 \approx -4 $,
which roughly corresponds to
$95\%$ CL exclusion.
Note that in the Bayesian  
framework the value $f_{\rm EDE}=0.1$ is excluded by around 
$3\sigma$, which is somewhat stronger than what we have obtained in the frequentist test here.
This suggests that our Bayesian constraints 
are affected by the prior-volume 
effects at a level of less than $1\sigma$.

\clearpage
\pagebreak 
\section{EDE away from the $\Lambda$CDM limit: Analysis with a high lower bound on $f_{\rm EDE}$}
\label{app:prior}

In this appendix, we present the results of an MCMC analysis of the \emph{Planck} 2018+BOSS+$S_8$ data with an aggressive lower bound for the $f_{\rm EDE}$ prior, $f_{\rm EDE}>0.04$. The purpose of this experiment is to clarify if the dearth of evidence for EDE in the CMB and LSS datasets is an artifact of the sampler failing to sufficiently explore parameter space, owing to the parameter degeneracies that emerge in the $\Lambda$CDM limit $f_{\rm EDE}\rightarrow 0$, as claimed in, e.g., \cite{Smith:2019ihp} and \cite{Niedermann:2020dwg}. If this were true, then one would expect that an MCMC analysis with $f_{\rm EDE}$ strictly bigger than a certain threshold value would lead to qualitatively different results, as the sampler would never approach the $\Lambda$CDM limit where parameter-space volume effects could emerge. We consider $f_{\rm EDE} = 0.04$ to be a reasonable lower threshold, given that the best-fit value in the fit to {\it Planck}+BOSS is $f_{\rm EDE} = 0.047$ (see Table~\ref{table:params-P18+BOSS}).

\begin{figure*}[h]
\centering
 \includegraphics[width=0.8\textwidth]{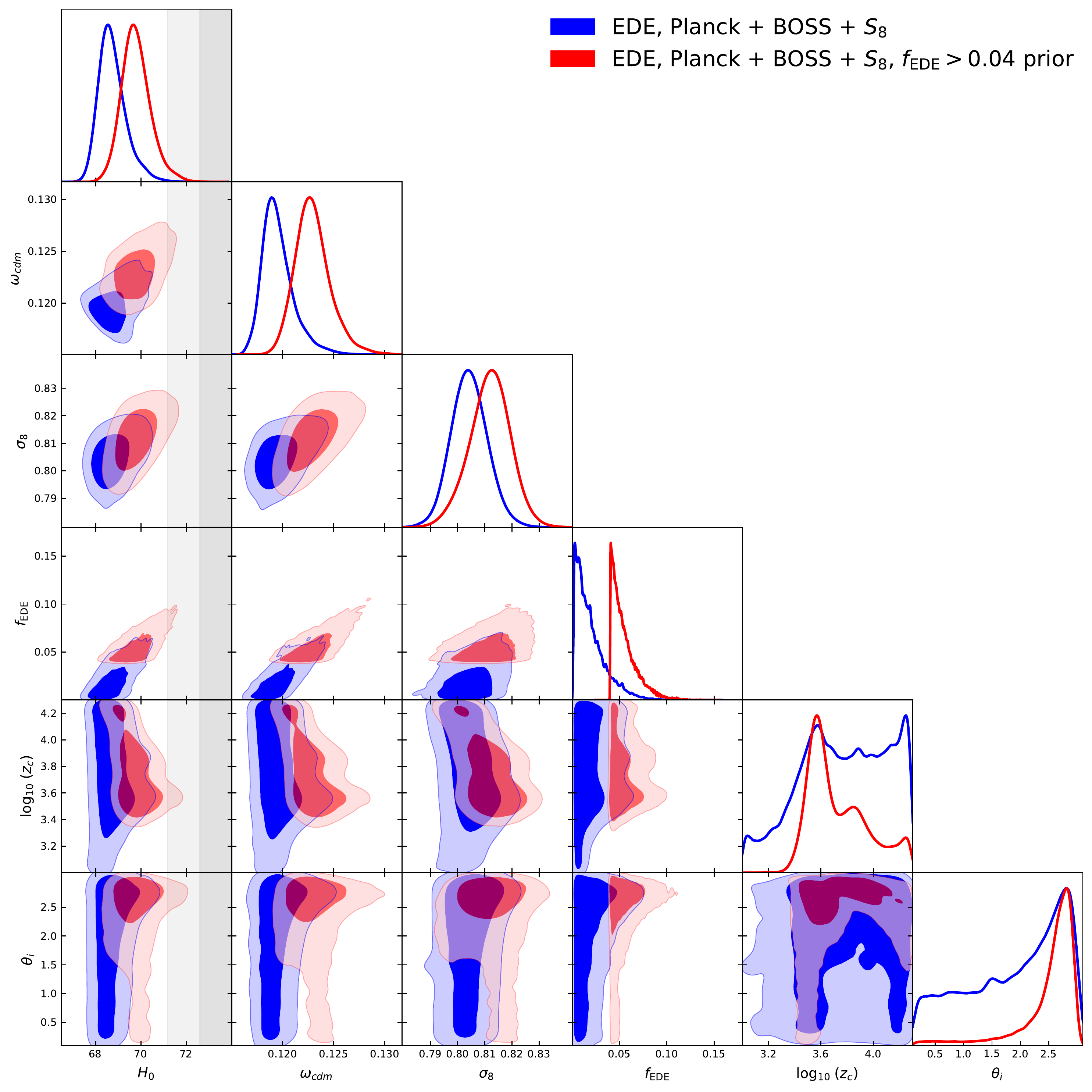}
    \caption{Cosmological parameter constraints from the \emph{Planck}+BOSS+$S_8$ dataset with two different priors on $f_{\rm EDE}$: $f_{\rm EDE}>0.001$ (blue) and $f_{\rm EDE}>0.04$ (red). The posterior for $f_{\rm EDE}$ exhibits substantial support on the lower bound of the prior in both cases, even the latter, for which the prior does not include the $\Lambda$CDM limit ($f_{\rm EDE} \approx 0$). This is consistent with our finding that there is no evidence for EDE in combined CMB and LSS data, and excludes parameter-space volume effects as an explanation for this result.
    The SH0ES $H_0$
    measurement is shown in gray.
    The dark-shaded 
and light-shaded 
contours mark 
$68\%$ and $95\%$ confidence intervals, respectively.
    }
    \label{fig:fEDE-prior}
\end{figure*}

The posterior distributions for the relevant cosmological parameters are shown in Fig.~\ref{fig:fEDE-prior}, and the marginalized limits are given in Table~\ref{table:params-fEDE-prior}. The posterior distribution for $f_{\rm EDE}$ is strongly peaked on the lower boundary of the prior, and we find $f_{\rm EDE} < 0.084$ at 95\% CL, consistent with the analyses of this work and of \cite{Hill:2020osr}, both of which find no evidence for EDE. This indicates that the cosmological constraints reported here, and in \cite{Hill:2020osr}, are not due to a failure of the Metropolis-Hastings algorithm to explore parameter space.

Finally, we perform the following frequentist test. We evaluate the {\it Planck}+BOSS+$S_8$ likelihood for the EDE model parameters fixed to the values given in Eq.~\eqref{hillparams} (which suffice to resolve the Hubble tension), and find the best-fit values of
all remaining nuisance parameters, to compute an effective $\chi^2$-statistic. The parameters in Eq.~\eqref{hillparams} result from an analysis of a combined data set that includes the SH0ES measurement, BAO, and other experiments, while neglecting LSS data from DES, KV-450, and HSC~\cite{Hill:2020osr}; this data set combination was chosen to match those used in the EDE analyses of~\cite{Poulin:2018cxd,Smith:2019ihp}. Of all the analyses and data set combinations considered in \cite{Hill:2020osr}, this came closest to consistency with the SH0ES measurement alone. One might imagine that these parameters reflect an underlying preference of the CMB and LSS data for non-zero $f_{\rm EDE}$, which is otherwise (i.e., in dataset combinations that do not include SH0ES) obscured due to the failure of the sampler as described above. To test this hypothesis, we compare the resulting $\chi^2_{\rm eff}$ with the one obtained from the best-fit parameters of the baseline analysis of the {\it Planck}+BOSS+$S_8$ data (see Table~\ref{table:params-P18+BOSS-S8}). This gives
\begin{equation}
    \chi^2_{\rm eff}(\text{true best-fit})
    - \chi^2_{\rm eff}(\text{cosmology from Eq.~\eqref{hillparams}}) =
   3163.72 - 3170.2=-6.48\,,
\end{equation}
demonstrating that the (nearly) $H_0$-tension-resolving parameter set in Eq.~\eqref{hillparams} is indeed a worse fit to the data than that found with a direct MCMC analysis. This indicates that our constraints are not driven by prior volume effects.

\begin{table*}[htb!]
EDE away from the $\Lambda$CDM limit \\
Constraints from \emph{Planck} 2018 data  + BOSS DR12  + $S_8$ from DES+KV-450+HSC\vspace{2pt} \\
  \centering
  \begin{tabular}{|l|c|c|}
    \hline\hline Parameter & $f_{\rm EDE}>0.04$  ~~&~~~$f_{\rm EDE}>0$ ~~~\\ \hline \hline

    {\boldmath$\ln(10^{10} A_\mathrm{s})$} & $3.042~(3.027)_{-0.015}^{+0.014}$ & $3.038 \, (3.034) \, \pm 0.014$ \\

    {\boldmath$n_\mathrm{s}$} & $0.9763\,(0.9742)_{-0.0052}^{+0.0061}$ & $0.9696 \, (0.9624)^{+0.0042}_{-0.0051}$ \\

    {\boldmath$100\theta_\mathrm{s}$} & $1.041945 \, (1.041966) \, \pm 0.00030 $ & $1.04178 \, (1.04176) \, \pm 0.00035$\\

    {\boldmath$\Omega_\mathrm{b} h^2$} & $0.02274 \, (0.02278) \,^{+0.00019}_{-0.00017}$ &  $0.02259 \, (0.022433)^{+0.00016}_{-0.00018}$ \\

    {\boldmath$\Omega_\mathrm{cdm} h^2$} & $ 0.1229\,(0.1219)_{-0.002}^{+0.0014}$ & $0.11958 \, (0.11951)^{+0.00096}_{-0.0018}$ \\

    {\boldmath$\tau_\mathrm{reio}$} & $0.05282\,(0.04781)_{-0.0072}^{+0.0074}$ & $0.0535 \, (0.0521)_{-0.0075}^{+0.0069}$\\

    {\boldmath$\mathrm{log}_{10}(z_c)$} & $3.746(3.67)_{-0.28}^{+0.17}$ & $3.77 \, (4.24)^{+0.51}_{-0.33}$ \\

    {\boldmath$f_\mathrm{EDE} $} & $<0.08384\,(0.04078)$ & $< 0.0526 \, (0.0115)$\\

    {\boldmath$\theta_i$} & $2.522(2.505)_{-0.064}^{+0.46}$ & $1.91(1.55)_{-0.47}^{+1.2} $\\

    \hline

    $H_0 \, [\mathrm{km/s/Mpc}]$ & $69.77\,(69.39)_{-0.72}^{+0.55}$ & $68.73 \, (67.92)^{+0.42}_{-0.69}$ \\

    $\Omega_\mathrm{m}$ & $0.3007~(0.3017)
    \pm 0.0052$ & $0.3024 \,(0.3091)\pm 0.0050$ \\

    $\sigma_8$ & $0.8115\,(0.8040)_{-0.0073}^{+0.008}$ & $0.8044 \, (0.8023)_{-0.0069}^{+0.0060}$ \\
    
    $S_8$ & $0.8126 \, (0.8063) \, \pm 0.0096$ & $0.8075 \, (0.8143)\, \pm 0.0092$ \\
    \hline
  \end{tabular} 
  \caption{
  The mean (best-fit) $\pm1\sigma$ constraints on the cosmological parameters in the EDE scenario with $n=3$, as inferred from the combination of BOSS FS+BAO, \emph{Planck} 2018 TT+TE+EE + low $\ell$ + lensing, and DES+KV-450+HSC data.  Upper and lower limits are quoted at 95\% CL. We present the results of two analyses differing by a lower prior bound on $f_{\rm EDE}$:
  baseline physical choice $f_{\rm EDE}>0$ (right column) and artificial unphysical choice $f_{\rm EDE}>0.04$ (left column).
  }
  \label{table:params-fEDE-prior}
\end{table*}

\clearpage

\section{EFT-based vs. Standard 
BOSS
likelihoods}
\label{app:fs8}

In this appendix, we present a joint analysis of the {\it Planck} 2018 data
and the ``consensus'' BOSS DR12 FS+BAO likelihood \cite{Alam:2016hwk}, and compare the results to those found in Sec.~\ref{sec:BossFS} for the joint analysis of {\it Planck} 2018 and our EFT-based BOSS likelihood. The consensus BOSS DR12 FS+BAO likelihood is obtained from a fit of the BAO and $f\sigma_8$ parameters to the BOSS data using a fixed power spectrum template, which was computed for a fiducial cosmology consistent with {\it Planck} $\Lambda$CDM. We use the \texttt{Monte Python}
implementation of this likelihood, after having corrected a non-negligible bug reported on GitHub\footnote{\url{https://github.com/brinckmann/montepython_public/issues/112}},
which had not been fixed in the official code distribution at the time when this manuscript was finalized.

\begin{figure*}[ht!]
\begin{center}
\includegraphics[width=0.8\textwidth]{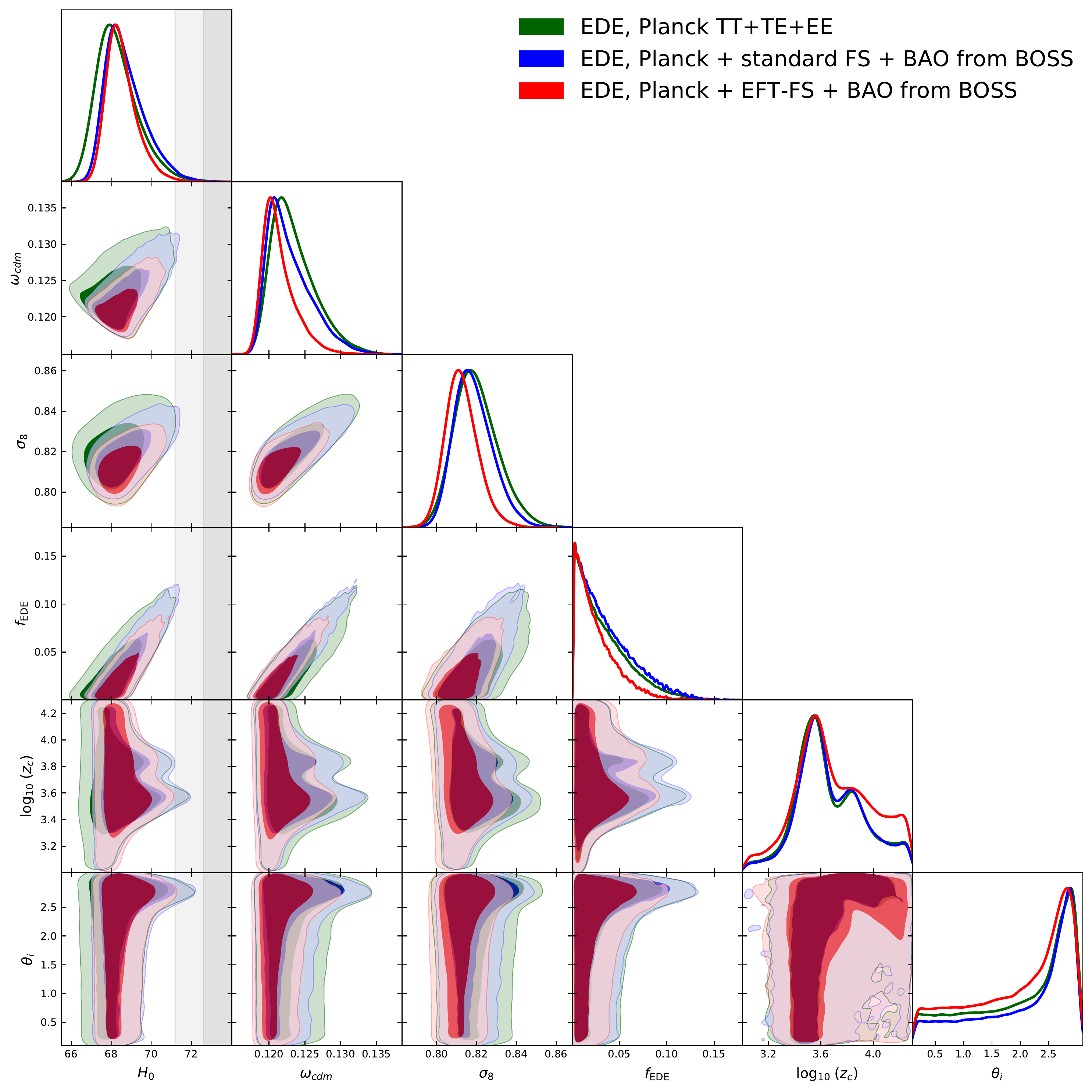}
\end{center}
\caption{
Posterior distributions for the
parameters extracted from the joint {\it Planck} 2018 TT+TE+EE+low $\ell$+lensing + BOSS FS+BAO data.
We show the results obtained
using the standard FS+BAO likelihood (in blue)
and the EFT-based likelihood (in red).
For reference, we also display the  
 constraints from the Planck 2018 primary CMB data alone (TT+TE+EE), obtained in \cite{Hill:2020osr}. 
The gray band shows the $H_0$ measurement from SH0ES, for comparison.
The dark-shaded 
and light-shaded 
contours mark 
$68\%$ and $95\%$ confidence intervals, respectively.
\label{fig:planck+BOSS-fs8} } 
\end{figure*}

The resulting 1D and 2D posterior distributions are shown in Fig.~\ref{fig:planck+BOSS-fs8}, along with those obtained in our baseline analysis using the complete EFT-based full-shape likelihood.
The corresponding  marginalized limits are presented in Table~\ref{table:params-P18+BOSS-eft_vs_stand}. One can see that the full likelihood 
yields a narrower $\omega_{cdm}$ posterior compared to the ``compressed'' standard likelihood, which is 
consistent
with the improvement expected from 
the power spectrum shape information.  
Thus, unlike the EFT-based likelihood, the standard BAO+RSD likelihood does not allow one 
to break the degeneracy between $f_{\rm EDE}$ and $\omega_{\rm cdm}$. This explains why the addition of the standard likelihood does not noticeably improve the primary CMB-only EDE constraints.

\begin{table*}[htb!]
Constraints from \emph{Planck} 2018 data  + BOSS DR12 for EDE ($n=3$) \vspace{2pt} \\
  \centering
  \begin{tabular}{|l|c|c|}
    \hline\hline Parameter &Standard
    BOSS likelihood~~&~~~EFT
    BOSS
    likelihood~~~\\ \hline \hline

    {\boldmath$\ln(10^{10} A_\mathrm{s})$} & $3.053\,(3.060)_{-0.016}^{+0.015}$ & $3.047 \, (3.049) \, \pm 0.014$ \\

    {\boldmath$n_\mathrm{s}$} & $0.9713\,(0.9790)_{-0.0085}^{+0.0055}$ & $0.9696 \, (0.9717)^{+0.0046}_{-0.0068}$ \\

    {\boldmath$100\theta_\mathrm{s}$} & $1.04185 \, (1.04200)_{-0.00032}^{+0.00035}$ & $1.04172 \, (1.04126) \, \pm 0.00032$\\

    {\boldmath$\Omega_\mathrm{b} h^2$} & $0.02256 \,(0.02269)_{-0.00022}^{+0.00018}$ &  $0.02255 \, (0.02245) \, \pm 0.00018$ \\

    {\boldmath$\Omega_\mathrm{cdm} h^2$} & $0.1230\,(0.1269)_{-0.004}^{+0.0018}$ & $0.1215 \, (0.1243)^{+0.0013}_{-0.0029}$ \\

    {\boldmath$\tau_\mathrm{reio}$} & $0.05688\,(0.05755)_{-0.0078}^{+0.0071}$  & $0.0553 \, (0.0543)_{-0.0075}^{+0.0069}$\\

    {\boldmath$\mathrm{log}_{10}(z_c)$} & $3.67\,(3.813)_{-0.27}^{+0.22}$ & $3.71 \, (3.52)^{+0.26}_{-0.33}$ \\

    {\boldmath$f_\mathrm{EDE} $} & $<0.096\,(0.076)$ & $< 0.072 \, (0.047)$\\

    {\boldmath$\theta_i$} & $2.122\,(2.934)_{-0.29}^{+0.98}$ & $2.023(2.734)_{-0.34}^{+1.1} $\\

    \hline

    $H_0 \, [\mathrm{km/s/Mpc}]$ & $68.71\,(69.66)_{-1.2}^{+0.69}$ & $68.54 \, (68.83)^{+0.52}_{-0.95}$ \\

    $\Omega_\mathrm{m}$ & $0.3097\,(0.3096)_{-0.0063}^{+0.0062}$ & $0.3082 \,(0.3120)_{-0.0057}^{+0.0056}$ \\

    $\sigma_8$ & $0.8187\,(0.8286)_{-0.011}^{+0.008}$ & $0.8127 \, (0.8195)_{-0.0091}^{+0.0072}$ \\
    
    $S_8$ & $0.8316 \, (0.8363) \, \pm 0.012$ & $0.8237 \, (0.8275)\, \pm 0.011$ \\
    \hline
  \end{tabular} 
  \caption{
  The mean (best-fit) $\pm1\sigma$ constraints on the cosmological parameters in the EDE scenario with $n=3$, as inferred from the combination of 
  \emph{Planck} 2018 TT+TE+EE+low $\ell$+lensing data with
  the standard (left column) 
  and EFT-based (right column)
  BOSS FS+BAO measurements. The upper limit on $f_{\rm EDE}$ is quoted at 95\% CL.}
  \label{table:params-P18+BOSS-eft_vs_stand}
\end{table*}

\end{document}